\documentclass[fleqn,usenatbib]{mnras}

\usepackage{newtxtext,newtxmath}

\usepackage[T1]{fontenc}

\DeclareRobustCommand{\VAN}[3]{#2}
\let\VANthebibliography\thebibliography
\def\thebibliography{\DeclareRobustCommand{\VAN}[3]{##3}\VANthebibliography}

\usepackage{graphicx}	
\usepackage{amsmath}	

\def \rvir{{$r_\mathrm{vir}$} }
\def \mvir{{$M_\mathrm{vir}$} }
\def \svir{{$\sigma_\mathrm{v,vir}$} }
\def \Svir{{$S_\mathrm{vir}$} }

\def \l'{{``}}
\def \r'{{''}}

\usepackage[dvipsnames]{xcolor}

\usepackage{makecell}
\usepackage{booktabs}
\usepackage{physics}

\title[Phase-space of galaxy clusters]{On the phase-space structure of galaxy clusters from cosmological simulations}

\author [Marini et al.]
{I. Marini$^{1,2,3,4}$\thanks{ilaria.marini@inaf.it}   A. Saro$^{1,2,3,4}$, S. Borgani$^{1,2,3,4}$, G. Murante$^{2,3}$, E. Rasia$^{2,3}$, K. Dolag$^5$,\\~\\
{\rm {\LARGE W. Lin$^6$, N.R. Napolitano$^6$, A. Ragagnin$^{2,3}$,  L. Tornatore$^{2,3}$, Y. Wang$^6$}}
    \\\\
    $^1$ Astronomy Unit, Department of Physics, University of Trieste, via Tiepolo 11, I-34131 Trieste, Italy\\
    $^2$ INAF-Osservatorio Astronomico di Trieste, via G. B. Tiepolo 11, I-34143 Trieste, Italy\\
    $^3$ IFPU - Institute for Fundamental Physics of the Universe, Via Beirut 2, 34014 Trieste, Italy\\  
    $^4$ INFN--Sezione di Trieste, Trieste,  Italy\\
    $^5$ Universitäts-Sternwarte München, Fakultät für Physik, LMU Munich, Scheinerstr. 1, 81679 München, Germany\\
    $^6$ School of Physics and Astronomy, Sun Yat-sen University, Zhuhai Campus, 2 Daxue Road, Xiangzhou District, Zhuhai 519082, China\\
}

\date{Accepted 2020 November 3. Received 2020 November 3; in original form 2020 July 9. In publication}

\pubyear{2020}

\begin{document}
\label{firstpage}
\pagerange{\pageref{firstpage}--\pageref{lastpage}}
\maketitle

\begin{abstract}
 Cosmological N-body simulations represent an excellent tool to study the formation and evolution of dark matter (DM) halos and the mechanisms that have originated the universal profile at the largest mass scales in the Universe. In particular, the combination of the velocity dispersion $\sigma_\mathrm{v}$ with the density $\rho$ can be used to define the pseudo-entropy $S(r)=\sigma_\mathrm{v}^2/\rho^{\,2/3}$, whose profile is well-described by a simple power-law $S\propto\,r^{\,\alpha}$. We analyze a set of cosmological hydrodynamical re-simulations of massive galaxy clusters and study the pseudo-entropy profiles as traced by different collisionless components in simulated galaxy clusters: DM, stars, and substructures. We analyze four sets of simulations, exploring different resolution and physics (N-body and full hydrodynamical simulations) to investigate convergence and the impact of baryons. We find that baryons significantly affect the inner region of pseudo-entropy profiles as traced by substructures, while DM particles profiles are characterized by an almost universal behavior, thus suggesting that the level of pseudo-entropy could represent a potential low-scatter mass-proxy. We compare observed and simulated pseudo-entropy profiles and find good agreement in both normalization and slope. We demonstrate, however, that the method used to derive observed pseudo-entropy profiles could introduce biases and underestimate the impact of mergers. Finally, we investigate the pseudo-entropy traced by the stars focusing our interest in the dynamical distinction between intracluster light (ICL) and the stars bound to the brightest cluster galaxy (BCG): the combination of these two pseudo-entropy profiles is well-described by a single power-law out to almost the entire cluster virial radius.
\end{abstract}




\section{Introduction}
    Galaxy clusters are associated with the collapse of the largest gravitationally bound overdensities in the initial density field of the Universe. Their abundance and their clustering properties are important cosmological probes that allow us to test the initial conditions of the Universe and to constrain the cosmological parameters \citep[e.g.][]{allen2011cosmological}. Their formation and evolution are driven by gravity-induced dynamics, while several baryonic processes (e.g., radiative cooling, star formation, and AGN feedback) play a major role in determining their observational properties at different wavelengths \citep{kravtsov2012formation}. The hierarchical assembly of clusters through the dynamical instability of dark matter (DM) dominated density perturbations should leave its imprint on the phase-space structure of these objects. In this context, cosmological numerical simulations are instrumental to describe in detail the phase-space structure of galaxy clusters and, ultimately, to capture the complexity of their formation process. In fact, cosmological simulations demonstrated that the equilibrium configuration of DM halos is characterized by a quasi-universal density profile at least out to the virial radius \cite[NFW density profile,][]{navarro1996structure,navarro1997universal}. 
    On the other hand, several studies \cite[e.g.,][]{taylor2001phase,dehnen2005dynamical} have shown that rather than to its density profile, cluster and DM halo formation and evolution might be more deeply connected to another, possibly more fundamental quantity: the ``pseudo-entropy" profile $S(r)$. This quantity is defined in terms of the velocity dispersion profile $\sigma_\mathrm{v}(r)$ and the density profile $\rho(r)$:
    \begin{equation}
        S(r)=\dfrac{\sigma_\mathrm{v}^2(r)}{\rho^{\,2/3} (r)}.
    \end{equation}
    The \l'phase-space density\r' $Q(r) = S^{-3/2}(r)$ is equivalently discussed in the literature.
    Empirically, it has been shown that $S(r)$ (or analogously $Q$) closely follows a power-law in radius in simulated galaxy-size \citep{taylor2001phase} and cluster-size halos \citep{rasia2004dynamical,ascasibar2004physical}, a result that has been confirmed by observations \citep[e.g.][]{biviano2013clash,biviano2016dynamics,capasso2019galaxy}. \citet{taylor2001phase} found from N-body simulations $Q\propto r^{-1.82}$ (corresponding to $S(r) \propto r^{\alpha}$ with $\alpha=1.21$). Similarly, \citet{rasia2004dynamical} derived $Q\propto r^{-1.85}$, i.e. $\alpha\simeq 1.23$, in agreement with the analysis of \citet{dehnen2005dynamical}. Moreover, the universality of halo density profiles can be recovered starting from the power-law behavior of the phase-space density profile and the Jeans equation, under the assumption of an isotropic, spherically symmetric equilibrium mass distribution \citep{dehnen2005dynamical}. This result motivates the study of pseudo-entropy as a quantity intrinsically connected to the process of halo formation. However, the underlying physical reason leading to the power-law dependence of pseudo-entropy is still unclear. 
    
    The power-law behavior of pseudo-entropy profiles in DM halos was also independently derived by \citet{faltenbacher2007entropy}, starting from the analogy with the entropy of the intracluster gas $S_X(r)$ which is generally defined as $S_X\propto T_g\, \rho_g^{-2/3}$, where $T_g$ is the gas temperature and $\rho_g$ is the gas density. Spherical gas accretion models predict gas entropy to scale with the clustercentric distance as $S_X\propto r^{1.1}$ \citep{tozzi2001evolution,voit2003origin}. Indeed outside the central region, mostly affected by non-gravitational processes \citep[e.g.][]{borgani2011cosmological}, the slope obtained in non--radiative hydrodynamical simulations agrees with the observed values (e.g. \citealt{voit2005baseline}) and gas and DM entropy profiles follow one another very closely. 
    
    In this paper, we analyze an extended set of cosmological hydrodynamical simulations to investigate the pseudo-entropy profiles of simulated galaxy clusters. For the first time, we present the pseudo-entropy profiles associated with different collisionless components in clusters, namely DM, stars in the main halo, as well as substructures whose dynamic is expected to trace that of galaxies within clusters. The possibility of combining simulations including different resolution and physics (i.e. N-body and hydrodynamical simulations with several baryonic effects) allows us to study in detail both numerical and dynamical/physical processes that determine the phase-space structure of galaxy clusters.    

    Stars in the main halo, which have been proven to be composed of two different dynamical populations, were further investigated. Several studies of both observational data and simulations have shown the existence of two different stellar components in the main halo of galaxy clusters.  A substantial fraction of these stars is confined within the brightest cluster galaxy (BCG). \citet{dubinski1998origin} investigated the origin of the BCG showing that close encounters and halo merging naturally produce a massive central galaxy with surface brightness and velocity dispersion profiles resembling those of the BCGs. The other fraction is not gravitationally bound to any particular galaxy and constitutes the so-called ``intracluster light" (ICL). The distribution of the ICL involves physical scales comparable to those over which the DM component is distributed \citep[e.g.][]{dubinski1998origin}, so it is reasonable to expect that this component traces the global gravitational potential of its hosting cluster \citep{montes2018intracluster}. Simulations predict that the ICL forms at relatively late times \citep[$z < 1$; e.g.,][]{contini2013formation,monaco2006diffuse,murante2007importance} and it is thought to arise primarily from the tidal stripping of stars from infalling groups and satellite galaxies during the hierarchical accretion of the cluster \citep[e.g.][]{murante2004diffuse}. We test the differences in their dynamical properties in relation to the pseudo-entropy profiles traced singularly by the two to disentangle the distinct contributions to the pseudo-entropy profile of all the stars. Indeed, in the hypothesis of a strong correlation between distinct formation mechanisms and phase-space structure, we expect to detect a corresponding difference in the pseudo-entropy profiles.
    
    In recent years, some observational analyses have deepened the study on the pseudo-entropy as traced by the hosted galaxies in galaxy clusters \citep{biviano2013clash,annunziatella2016clash,capasso2019galaxy}. These results have shown the existence of the power-law feature, also for this tracer. We plan to provide a computational counterpart in that regard: the dynamics of real galaxies is expected to be traced by the substructures within the simulated halos. 
    
    This paper is organized as follows: in Sec. \ref{sec-sim} we briefly describe the details of the simulation setup. In Sec. \ref{sec-method} we discuss the universality of the pseudo-entropy profiles. Sec. \ref{sec:S-DM} presents our results for the properties of the pseudo-entropy profiles associated with DM particles. In Sec. \ref{sec:S-sub}, we discuss the properties of the pseudo-entropy profiles associated with substructures while the pseudo-entropy profiles associated with stars (including the intracluster light component, ICL) are shown in Sec. \ref{sec:S-stars}. Sec. \ref{sec:data} presents a comparison with observational results on the phase-space density traced by the total matter and galaxies within a real cluster sample. Finally, Sec. \ref{sec-summary} summarizes our main results.

    \section{Simulation details}
    \label{sec-sim}
   \begin{table}
        \caption{A summary of the main characteristics of the analyzed simulations. For each set of simulations we report: DM and gas particles mass; DM, gas, star particles and black hole (Plummer-equivalent) softening lengths at redshift $z=0$; and the total number of clusters analyzed.}
        \centering
        \resizebox{0.47\textwidth}{!}{
            \setlength{\tabcolsep}{8pt}
            \begin{tabular}{c c c c c}
                \\
                & \thead{Hydro-1x} & \thead{Hydro-10x} & \thead{DM-10x} & \thead{DM-100x} \\\\
                \hline \\
                $M_\mathrm{DM}\, [10^8h^{-1}M_{\odot}]$ & $8.3$ &$0.83$ &$1.00$ &$0.10$ \\\\
                $M_{\mathrm{gas}}\, [10^8h^{-1}M_{\odot}]$&$3.3$ & $0.33$& - & - \\\\
                $\epsilon_{DM}\,[kpc\, h^{-1}]$  &$3.75$ &$1.4$ &$1.4$ &$0.6$\\\\
                $\epsilon_{gas}\,[kpc\, h^{-1}]$ & $3.75$&$0.375$ & - & -\\\\
                $\epsilon_{star}\,[kpc\, h^{-1}]$ & $1.0$&$0.35$ & - & -\\\\
                $\epsilon_{BH}\,[kpc\, h^{-1}]$ & $1.0$&$0.35$ & - & -\\\\
                N Clusters & $29$ & $11$ & $29$ & $12$\\\\
                \hline
            \end{tabular}
            \label{tab:simulation_value}}
    \end{table}
    Simulations were performed with the code GADGET-3, an improved version of the Tree/PM Smoothed-Particle-Hydrodynamics (SPH) public code GADGET-2 \citep{springel2005cosmological}. We analyze a set of DM-only simulations (DM) and a set of hydrodynamical simulations (Hydro). For each set, we carried out simulations at two different levels of resolution (see Table \ref{tab:simulation_value}). At the base resolution (1x hereafter), for the Hydro set we adopt a DM mass particle of $8.3\times10^8 h^{-1}M_{\odot}$ and an initial mass of the gas particle of $3.3\times10^8 h^{-1}M_{\odot}$. At intermediate (10x hereafter) resolution, we have both DM-only and hydrodynamical simulations, while at high resolution (100x hereafter) we include a set of DM-only simulations.
    The characteristics of the four sets of simulations are summarized in Table \ref{tab:simulation_value}.
    
    The set of simulated clusters, named Dianoga \citep[][and references therein]{bassini2020DIANOGA}, were extracted from a parent N-body box of size $1\, h^{-1}$ Gpc and resimulated adopting the zoom-in technique as implemented by \cite{tormen1996adding}. The adopted cosmology is a $\Lambda$CDM model with $\Omega_M=0.24$, $\Omega_b=0.037$ for the total matter and baryon density parameters, $n_s=0.96$ for the slope of the primordial power spectrum, $\sigma_8 =0.8$ for the normalization of the power spectrum, $h_0=0.72$ for the Hubble parameter in units of 100 km s$^{-1}$Mpc$^{-1}$. At the base resolution, each simulation describes the evolution of a Lagrangian region centered on the $24$ most massive clusters in the initial simulated box, all having mass $M_{200}\geq 8 \times10^{14}h^{-1} M_{\odot}$ and $5$ isolated smaller clusters with $M_{200}$ in the range $(1-4) \times10^{14}h^{-1} M_{\odot}$; however, for the other resolutions, the number of clusters employed is different and varies for each simulation (we report the exact numbers in the last row of Table \ref{tab:simulation_value}).

    The version of the GADGET-3 code used for the hydrodynamical simulations presented here includes a higher-order kernel function for the SPH interpolation, a time-dependent artificial viscosity, and artificial conduction as described by \cite{beck2015improved}, to which we refer for further details. The sub-resolution model for star formation and galactic outflows driven by SN feedback are implemented according to the original model by \citet{springel2003cosmological} while metal enrichment and chemical evolution, whose stellar yields are specified in \cite{biffi2017history,biffi2018enrichment,biffi2018origin}, follow the formulation described in \cite{tornatore2007chemical}. In addition, the AGN feedback model is implemented as outlined in Appendix A of \cite{ragone2013brightest} with a new prescription for the coupling of the AGN feedback energy to the gas particles \citep{planelles2013baryon,planelles2013role}. We note that the set of Hydro-1x simulations is the one originally presented in \cite{ragone2018bcg}, while the set of Hydro-10x has been presented in \cite{bassini2020DIANOGA}.

    \subsection{Identifying substructures and stellar dynamical components}
    \label{subsec:subfind}
    To identify precisely locally overdense and self-bound particle groups distinct from the main structure, we run the SubFind algorithm \citep{springel2001populating,dolag2009substructures} on catalogs of groups of particles identified by a Friend-of-Friend (FoF) algorithm with a linking length of $b=0.16$ in units of the mean inter-particle separation. We assume a substructure to be resolved if it includes a minimum of 50 (DM or stellar) particles. 
   
    Although this procedure works well for identifying substructures in a simulation with different particle species, it does not split the stellar population into the diffuse ICL and the stars bound to the BCG. In order to identify the two stellar components within our simulations, we employed a modified version of SubFind \citep{dolag2010dynamical} which resorts to a criterion of dynamical segregation of BCG and ICL stars to separate such two components. The algorithm starts by fitting the velocity distribution of all the stars with a double Maxwellian; each single Maxwellian distribution is assumed to correspond to one of the two distinct stellar components: namely the ICL, associated with the distribution with the larger velocity dispersion, and the BCG with the smaller one. Then, the algorithm computes the gravitational potential contributed by the star particles contained within a sphere of a given fiducial radius, centered on the center of the halo. The radius of the sphere is initially assumed to be a fraction of the virial radius and divides the star particles into two components; for each subgroup, the fitting procedure of the velocity distribution is performed again with a single Maxwellian. By varying the radius and recomputing the gravitational potential, the procedure is repeated until the velocity distributions of the two components converge to the two velocity distributions inferred from the original global fit. This last step unbinds the star particles in the two components. Unlike in the observational studies whereby the ICL is identified by projected surface brightness criteria, this method provides a more physically motivated result using the full six-dimensional phase-space information, although the resulting ICL cannot be directly compared to that obtained from observations. 
   
    \subsection{Dynamical state}
    \label{subsec:dyn}
    Estimates of the dynamical state of single clusters became important when investigating the impact of the internal equilibrium of these systems in relation to their pseudo-entropy profile. We classify only the clusters in the Hydro-1x sample. This is performed following the prescription described in \citet{biffi2016nature} to which we refer for further details.
   
   The method is based on two properties: the center shift (identified as the distance between the position of the minimum of the gravitational potential $\mathbf{x_{min}}$ and the center of mass $\mathbf{x_{cm}}$) and the fraction of mass in substructures $f_\mathrm{sub}$. A halo is classified as relaxed if both the following conditions are satisfied:
    \begin{equation}\label{eq:biff+2016}
          \begin{cases}
             \delta r = || \mathbf{x_{min}} - \mathbf{x_{cm}} ||/r_{200}<0.07\\ \\
              f_\mathrm{sub} = \dfrac{M_{TOT,sub}}{M_{TOT}} <0.1
          \end{cases}       
    \end{equation}
    where $M_{TOT}$ is the total mass and $M_{TOT,sub}$ is the total mass in substructures. If neither is satisfied, then the cluster is classified as disturbed, while it is tagged as partially disturbed if only one of the above two criteria is not satisfied. After applying this classification on the 29 main clusters of the Hydro-1x set at redshift $z=0$, we find  6 relaxed, 8 disturbed systems, and 15 intermediate cases.

    \section{Self-similarity of pseudo-entropy profiles}
    \label{sec-method}
    \citet{ludlow2010secondary} found that the pseudo-entropy profiles may not hold the power-law behavior when approaching the virial radius due to the proximity of the last accreted shell. Our aim is to broaden the analysis of the universality of the pseudo-entropy profiles traced by different collisionless components in the simulated clusters. 
    
    Our analysis is extended out to redshift $z=2$, which is the epoch when massive clusters assemble. All the particle positions and velocities are computed in the rest-frame of the cluster center, which is identified as the particle within the central FoF group or the main halo having the minimum value of the gravitational potential. Cluster radii are defined in units of the virial radius $r_\mathrm{vir}$, in order to better capture the universal behavior of the self-similar scaling of the simulated clusters. The virial radius of a halo at redshift $z$ is defined as the radius encompassing a mean halo density of $\Delta_\mathrm{vir}(r)\rho_c(z)$, where $\rho_c(z)$ is the critical cosmic density at redshift $z$ and $\Delta_\mathrm{vir}$ is the redshift-dependent virial overdensity predicted by spherical-collapse for a given cosmological model \citep[e.g.][]{bryan1998statistical,eke1996cluster}\footnote{In a similar way, we can define $r_\Delta$ as the radius encompassing a mean overdensity equal to $\Delta$ times the critical density of the universe at that redshift $\rho_c(z)$.}. We provide here below a short description of the scaling associated with the velocity dispersion profiles, and consequently to the pseudo-entropy profiles, while the full derivation is provided in the Appendix \ref{appendixA}.  
    
    In the self-similar model, particles within a sphere of radius \rvir at redshift $z$ have a measured velocity dispersion profile $\widetilde{\sigma_{\mathrm{v}}}(r,z)$ that scales as a function of the virial radius $r_\mathrm{vir}$ (i.e. halo mass) and redshift
        \begin{equation}\label{eq:sigma(z)}
        \sigma_{\mathrm{v}} (r,z) =\frac{\widetilde{\sigma_{\mathrm{v}}}(r,z)}{r_{\mathrm{vir}}} \, \left[\dfrac{\Delta_{\mathrm{vir}}(z)}{\Delta_{\mathrm{vir}}(0)}\,E^2(z)\right]^{-1/2},
    \end{equation}
    where $E(z) = [\Omega_M (1+z)^3+\Omega_{\Lambda}]^{1/2}$ provides the redshift dependence of the Hubble parameter: $H(z)=E(z)H_0$. From this relation, we derive the scaling on the measured pseudo-entropy $\widetilde{S}(r,z)$:
    \begin{equation}\label{eq:S(r,z)}
        S(r,z) = \dfrac{\widetilde{S}(r,z) }{r^2_{\mathrm{vir}}\, E^{\,2/3}(z)\left(\dfrac{\Delta_{\mathrm{vir}}(z)}{\Delta_{\mathrm{vir}}(0)}\right)^{1/3}}.
    \end{equation}

\begin{figure}
        \includegraphics[width=7.2cm,scale=0.5,angle=0.0]{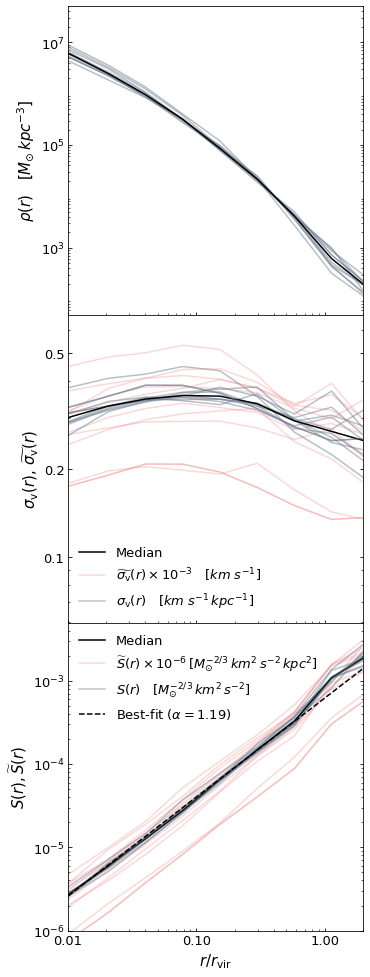}
        \caption{From the top panel: density, velocity dispersion, and pseudo-entropy profiles as traced by the DM particles in the Hydro-10x at $z=0$. The colored lines report the profiles traced by each cluster while the black is the median. In the middle panel, the grey and pink lines represent the velocity dispersion profiles $\sigma_\mathrm{v}(r)$  and  $\widetilde{\sigma_\mathrm{v}}(r)$ of each cluster respectively scaled and not-scaled according to Eq. \ref{eq:sigma(z)}. The black lines trace the median profiles. Similarly, in the bottom panel, the grey and pink lines represent the scaled and not-scaled profiles of the pseudo-entropy $S(r)$ according to Eq. \ref{eq:S(r,z)}, the black solid line illustrates the median trend while the dashed one is the best-fit.
        }
        \label{fig:Fake_Faltenbacher-bis}
    \end{figure}

    We show in Fig. \ref{fig:Fake_Faltenbacher-bis} how the rescaling proposed in Eq. \ref{eq:sigma(z)} and \ref{eq:S(r,z)} effectively captures the universal behavior of the individual pseudo-entropy profiles. More in detail, we illustrate the radial profiles of density, velocity dispersion, and pseudo-entropy, as traced by DM particles in the main halo given by Subfind within $2$ \rvir for the entire cluster sample of the Hydro-10x simulation at $z=0$. Individual cluster density profiles are shown as thin grey lines, while the solid black line represents the median profile. We note that no further scaling needs to be applied to the density profiles (upper panel) once the radial distance is expressed in terms of the virial radius. On the other hand, simply rescaling radii in units of the virial radius is not sufficient to properly capture the universality of the other two phase-space quantities (velocity dispersion $\widetilde{\sigma_{\mathrm{v}}}$ and pseudo-entropy $\widetilde{S}$ profiles), shown in pink thin lines in the central and bottom panels. For velocity dispersion and pseudo-entropy profiles, universality is recovered once such quantities are expressed in terms of $\sigma_{\mathrm{v}}$ and $S$, or in other words, are rescaled according to Eq. \ref{eq:sigma(z)} and \ref{eq:S(r,z)} (thin grey lines).
    
    Fig. \ref{fig:Fake_Faltenbacher-bis} confirms the remarkable power-law shape of the pseudo-entropy profiles (dashed black line) of the DM component, which is stable from the innermost resolved radius out to nearly the virial radius, in simulations including hydrodynamics and baryonic physics. As discussed in \citet{ludlow2010secondary}, the outer region is most likely associated with the transition from the inner, relaxed parts, to the dynamically more active outer parts, where infalling material has not yet had time to undergo phase-mixing and relaxation. Such an upturn is present also in the self-similar solution of \citet{bertschinger1985self} and it might be a general feature of the outer pseudo-entropy profiles of DM halos. On the other hand, the density profile corresponding to the Bertschinger solution (a power-law with constant slope) differs significantly from the density profiles of DM halos as shown in the top panel of Fig. \ref{fig:Fake_Faltenbacher-bis}, which are better described by NFW profiles, whereby the logarithmic slope smoothly changes from --1 in the central regions to --3 in the outer regions. Similarly, in the central panel, we show velocity dispersion profiles, which are also clearly showing departures from a scale-free behavior. As the main driver of the mechanisms involved in halo formation (phase-mixing and violent relaxation) is gravity (which has a scale-free behavior), it is reasonable to expect that closely associated phase-space density quantities retain a scale-invariant behavior.
    A simple power-law suggests, therefore, the possibility of interpreting the pseudo-entropy as a key quantity in structure formation, thanks to its power-law behavior that provides a more fundamental dynamical attractor than either the velocity dispersion or density profile which individually do not have a power-law trend (surprisingly given the fact that it is derived from the combination of the density and the velocity dispersion profiles which, we stress, are not power-laws if taken singularly).

    In the following sections, we present our results on the simulated clusters in more detail, focusing on the different tracers of the phase-space, namely DM particles, substructures, stars belonging to the BCG and the diffuse stellar component surrounding the BCG. 

    \section{Pseudo-entropy profiles traced by DM}
    \label{sec:S-DM}
    
    Having motivated why the pseudo-entropy  profile traced by DM particles is considered as a fundamental diagnostic for the description of halo formation, we now investigate its behavior for simulations with different resolutions and including the description of different physical processes as well as studying its evolution. This will allow us to assess the robustness of its shape against both numerical resolution and physical processes that add to gravitational instability. 
    
    \subsection{The effect of resolution}
    
\begin{figure}
        \includegraphics[scale=0.6,angle=0.0]{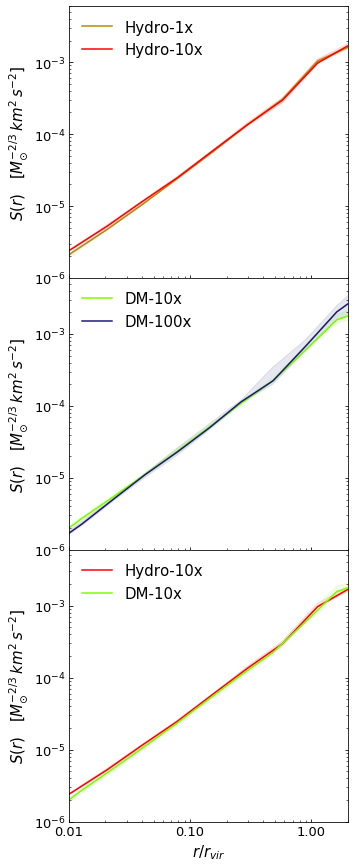}
        \caption{Comparison of the pseudo-entropy profiles traced by DM particles at redshift $z=0$ to test the convergence among the distinct simulations. In the top panel, we compare the median pseudo-entropy profiles obtained from the clusters in the Hydro-1x (in ocher) and Hydro-10x (in red) runs. In the central panel, we show the median pseudo-entropy profiles obtained from the DM-10x (in green) and DM-100x (in blue) simulations. The bottom panel reports the median pseudo-entropy profiles from the simulated clusters in the Hydro-10x (in red) and DM-10x (in green). For all the curves we plot the 68-th percentile region given by the cluster samples as a shaded area. 
         }
        \label{fig:PED-DM-sim}
    \end{figure}

    As shown in Fig. \ref{fig:Fake_Faltenbacher-bis}, DM particles in simulated clusters distribute in phase-space in such a way to predict a tight power-law shape of the pseudo-entropy profile. Fig. \ref{fig:PED-DM-sim} presents the pseudo-entropy profiles traced by the DM particles in the stack of clusters common to the four sets of simulations, all at $z=0$. The upper and the central panels compare each a pair of simulation sets including the same physics, but with different resolution: fully hydrodynamics and DM-only simulations, respectively. The bottom panel compares instead DM-only and Hydro simulations at the same resolution. Despite resolving structures with different sensitivity, due to the different mass resolutions and the presence of baryons, the phase-space robustly describes a similar power-law pseudo-entropy profile in all cases. We report the median profiles (solid lines) and the 68 percent dispersion (shaded areas) given by the cluster samples. The profiles are all in excellent agreement with each other, with small differences only in the innermost regions, where the effects of both resolution and baryonic processes become more relevant. 

    We then quantify the dependence of the pseudo-entropy profiles, traced by DM particles, on the resolution by computing the normalization $S_0$ and the logarithmic slope $\alpha$ for the DM-only and Hydro runs. We assume a power-law behavior as:
     \begin{equation} \label{eq:S(r)}
        S(r)= S_0 \,\left(\frac{r}{0.5\, r_{\mathrm{vir}}}\right)^{\alpha}.
    \end{equation}
    The fitting procedure is carried out with \textit{emcee} \citep{foreman2013emcee}, a Python implementation of the affine-invariant ensemble sampler for Markov Chain Monte Carlo. We use the median profile and the associated error to fit the profiles. Results are reported for the four sets of simulations in Table \ref{tab:DM}. We find that results on the normalization and slope for both the DM-only and the Hydro runs are consistent within 2$\sigma$ for the different resolutions, thus guaranteeing the convergence of our results against the resolution.

    \begin{table}
    \caption{Best-fit results for the normalization $S_0$ and exponent $\alpha$ in the power-law expression of Eq. \ref{eq:S(r)} for the median pseudo-entropy profiles at $z=0$ traced by DM particles.}
    \centering
        \begin{tabular}{  c c c   }
            \hline
            & \thead{$S_0$} & $\alpha$   \\\midrule
            \thead{Hydro-1x} & $\left(2.79^{+0.04}_{-0.04}\right) \times 10^{-4} $ & $1.20^{+0.01}_{-0.01}$  \\
            \\
            \thead{Hydro-10x} & $\left(2.72^{+0.05}_{-0.05}\right) \times 10^{-4} $ & $1.19^{+0.01}_{-0.01}$ \\
            \\
            \thead{DM-10x} & $\left(2.32^{+0.02}_{-0.02}\right) \times 10^{-4} $  & $1.23^{+0.01}_{-0.01}$ \\
            \\
            \thead{DM-100x} & $\left(2.39^{+0.06}_{-0.06}\right) \times 10^{-4} $  &$1.25^{+0.01}_{-0.01}$\\
            \\
            \hline
        \end{tabular}%
        \label{tab:DM}
    \end{table}
    
    \subsection{The impact of baryons}

    The bottom panel of Fig. \ref{fig:PED-DM-sim} shows the profiles for DM-only (in green) and Hydro (in red) simulations at the same resolution. Pseudo-entropy profiles derived from the DM particles in the Hydro runs have been rescaled by $(1-\Omega_b/\Omega_M)^{2/3}$ to properly compare them to their DM-only counterparts. 
    After this correction, we observe the two profiles to be almost indistinguishable over the entire radial range, besides the core regions, where the impact of baryons mostly influences the distribution of the DM particles. A new fit of the Hydro simulations, considering the aforementioned correction factor, gives values consistent with the DM-only results: for the Hydro-1x $S_0=2.48^{+0.04}_{-0.04}$ and for the Hydro-10x $S_0=2.42^{+0.04}_{-0.04}$. We note that the logarithmic slope $\alpha$ of the Hydro run profiles is slightly shallower ($\simeq 3 \sigma$) than its DM-only counterpart. Thus, the emerging picture of the pseudo-entropy profiles traced by the DM particles in the simulations agrees with the general result of a power-law with a fixed slope, a result that is supported against numerical resolution, indicating that this is a key quantity in the description of the gravity-driven collapse of non-linear structures. 
    
    \subsection{Evolution in redshift}
\begin{figure}
        \includegraphics[width=8.0cm]{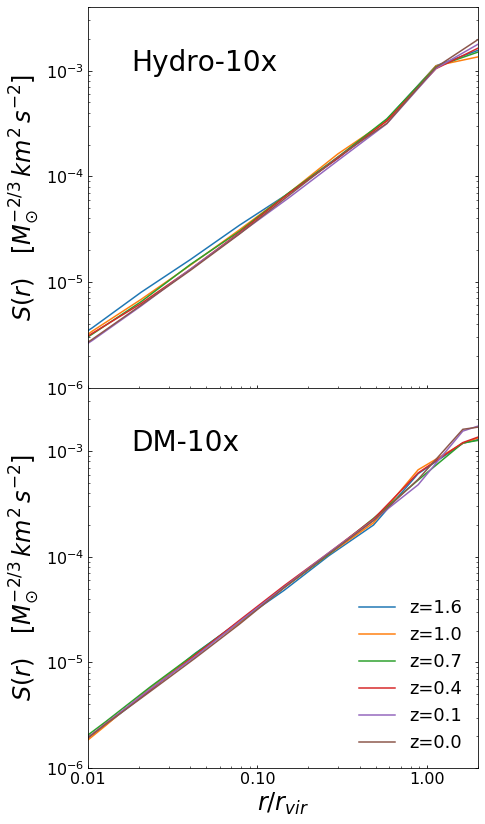}
        \caption{Top panel: median of the pseudo-entropy profiles traced by DM particles at different redshifts (as indicated in the legend in the bottom panel) for the clusters in the Hydro-10x simulation. Bottom panel: median of the pseudo-entropy profiles traced by DM particles at different redshifts for the clusters in the DM-10x simulation.}
        \label{fig:PED-DM-z}
    \end{figure}

    Fig. \ref{fig:PED-DM-z} describes the redshift evolution of the pseudo-entropy profiles. The top panel shows the median profiles traced by DM particles in the Hydro-10x runs for 6 different redshifts within $0\leq z\leq 1.6$ . The bottom panel illustrates the median profiles when traced by DM particles in the DM-10x. To not overload the plot we do not show the associated 68-th percentile regions (which we verify do not change significantly as a function of redshift). In both cases, the close similarity of the profiles highlights the self-similar scaling of the redshift evolution of pseudo-entropy profiles, as discussed in Sec. \ref{sec-method}. In particular, we note that the profiles are well-described by a power-law behavior, within the considered redshift range.
    
    The hydrodynamical run presents some tension in the innermost regions against the pure self-similar behavior in the DM-only case. These small deviations emerge for clusters at early times, because they tend to present higher entropy profiles with respect to their lower-redshift descendant. This reflects in a resulting systematic trend of decreasing in the slope $\alpha$ as a function of redshift.

    \subsection{The \mvir--\Svir relation}
    
    Simple scaling relations between basic cluster properties, such as the total virial mass \mvir and the dispersion velocity within the virial radius \svir \citep{bryan1998statistical,borgani1999velocity,evrard2008virial,munari2013relation,saro2013toward}, are naturally predicted by the self-similar model \citep{kaiser1986evolution,kaiser1992evolution}. From the analysis of an extended set of N-body simulations of galaxy clusters, \citet{evrard2008virial} found that massive DM halos closely adhere to the relation:
    \begin{equation}\label{eq:sigma_vir}
        \sigma_\mathrm{v,vir}(M_\mathrm{vir},z) = \sigma_\mathrm{v,15}\left(\frac{E(z) M_\mathrm{vir}}{10^{15} h^{-1}M_{\odot}}\right)^{\gamma}
    \end{equation}
    with a remarkably modest scatter $\sigma_{\sigma_\mathrm{v,vir}|M_\mathrm{vir}}\simeq0.04$, where $\sigma _\mathrm{v,15}=1082.9\pm 4.0\ \mathrm{km}\,\ \mathrm{s}\,^{-1}$ is the normalization at mass $10^{15}\ h^{-1}\ M_{\odot }$ and $\gamma=0.3361\pm 0.0026$ is the logarithmic slope, found to be within the virial expectation $\gamma=1/3$ considering the associated uncertainty. The tight scatter in this relation, in fact, makes \svir a rather accurate mass proxy. The best-fit scaling for our \mvir-- \svir relation is quite close to the virial expectation: we obtain $\gamma=0.347\pm0.013$ and an intrinsic logarithmic scatter of $\sigma_{\sigma_\mathrm{v,vir}|M_\mathrm{vir}}=0.048\pm0.007$, corresponding to a fractional uncertainty in mass at fixed observable of $\sigma_{M_\mathrm{vir}|\sigma_\mathrm{v,vir}} = 0.132\pm0.003$. 
    
\begin{figure}
        \includegraphics[width=8.2cm,scale=0.7,angle=0.0]{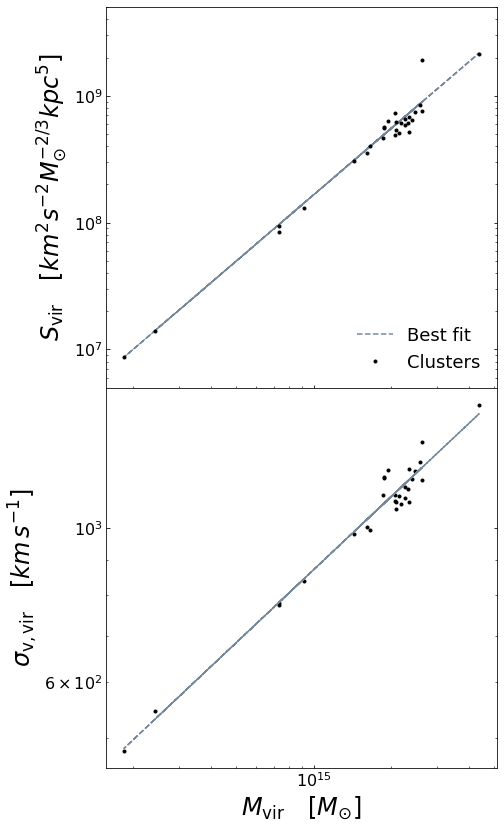}
        \caption{Scaling relation of the integrated pseudo-entropy (upper panel) and of the virial velocity dispersion (lower panel), both computed over all the DM particles within $r_\mathrm{vir}$, as function of the virial mass of the clusters in the Hydro-1x simulation at $z=0$. The dashed grey lines are the best-fit profiles for the scaling relations. 
        The second most massive halo in the sample is marked with a black cross to signal an outlier in the distribution of the integrated pseudo-entropy profile, as pointed out in Sec. \ref{subsec:outliers}.}
        \label{fig:scaling}
    \end{figure}

    Given the strong similarity of pseudo-entropy profiles, we can argue whether pseudo-entropy computed within the virial radius, could also provide an accurate, low-scatter halo mass proxy. 
    To this purpose, we define $S_{\mathrm{vir}}$ to be the integrated pseudo-entropy enclosed within the virial radius \rvir:
    \begin{equation}\label{eq:svir}
        S_{\mathrm{vir}} = 4\pi \int^{r_{\mathrm{vir}}}_0 \widetilde{S}\mathrm(r)\, r^2 \differential r.
    \end{equation}
   We assume the following scaling of the integrated pseudo-entropy with virial mass:
    \begin{equation}
        S_\mathrm{vir} (M_\mathrm{vir},z) = S_\mathrm{15} \left(\dfrac{E(z)M_\mathrm{vir}}{10^{15}h^{-1}M_{\odot}}\right)^{\gamma'}\,.
    \end{equation}
    We study DM particles in clusters from the Hydro-1x simulation at $z=0$ and compare the results obtained computing \Svir and the 3D velocity dispersion \svir as reported in the top and bottom panel of Fig. \ref{fig:scaling} respectively. For the \mvir-- \Svir relation, we find the slope $\gamma'= 1.74\pm 0.05$ with an intrinsic logarithmic scatter of $\sigma_{S_\mathrm{vir}|M_\mathrm{vir}}=0.20\pm0.03$, therefore corresponding to a fractional uncertainty in mass at fixed observable \Svir equal to $\sigma_{M_\mathrm{vir}|S_\mathrm{vir}} = 0.11\pm0.02$. This is a strong indication that the integrated pseudo-entropy might be a better mass proxy than the velocity dispersion, because its scatter against halo mass is even smaller than that of \svir. This would be of  particular relevance because, to estimate both quantities, one needs the same information on the cluster dynamics. However, to fully verify this claim, one requires a larger sample of clusters over a reasonably wider mass range to enhance the statistics. Moreover, observationally speaking, the integrated pseudo-entropy is recovered from the combination of two quantities which both carry their own uncertainties, thus one might expect to have an increase in the internal distribution due to the associated observational scatter. 

    \begin{figure*}

           \includegraphics[width=17.5cm,angle=0.0]{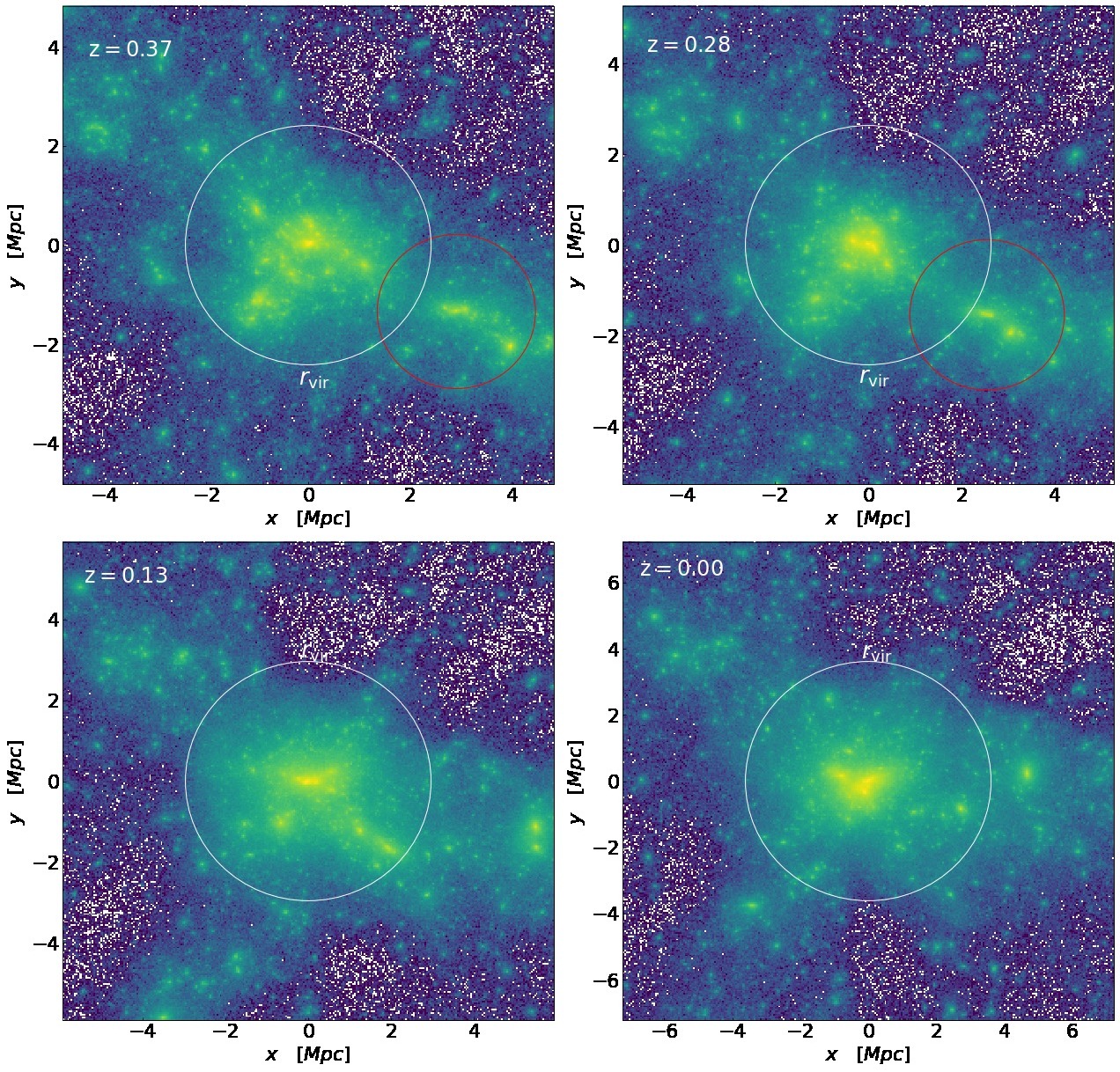} 
            \caption{The evolution of the density maps in logarithmic scale traced by the DM particles in the second most massive cluster reported in the top panel of Fig. \ref{fig:scaling} which appears as an outlier from the $S_{\mathrm{vir}}-M_{vir}$ scaling relation. The region is centered at the cluster center and spans a square of 4 times the virial radius of the main halo. The white circle marks the virial radius. The red smaller circle marks the trajectory of the second most massive halo in the region and its growth given that the radius of the circle is equal to its virial radius (as provided by SubFind). The orbiting object at redshift $z\simeq 0.13$ crosses the virial radius to merge with the main halo by redshift $z=0$.}
            \label{fig:D28}
    \end{figure*}

    \subsection{Outliers}
    \label{subsec:outliers}
    Self-similarity can be broken down if the scale-free evolution of a halo is distressed. In this event, one expects the disturbed cluster to not follow the scaling relation, but rather to be an outlier in the overall distribution. To this end, we note the presence of an obvious outlier in the \mvir--\Svir relation shown in Fig. \ref{fig:scaling} represented by the second most massive cluster which we signal in the plot with a black cross. To understand the nature of the outlier, we show in Fig. \ref{fig:D28} the recent evolution of its density maps. The four panels illustrate the density maps traced by DM particles within a region centered on the cluster center with size $4\,r_\mathrm{vir}$, at four redshifts. Brighter colors indicate higher densities. The white circle marks the virial radius. The red smaller circle indicates a second halo which is falling into the cluster potential and reaching the central regions at around $z=0.13$, as displayed in the left bottom panel. The size of the circle is equal to the virial radius of this second halo, as provided by SubFind. At this late redshift, the second halo has crossed the volume enclosed by the virial radius of the main halo and it is being incorporated. This merging process is completed by $z=0$. The mass ratio 1:5 is fairly large, thus the recent merging represents an event that strongly impacts the dynamical equilibrium of the main halo. Indeed, the sudden change in the internal dynamics is reflected in the pseudo-entropy, which significantly increases at late times, while the system has not yet had the time to virialize and settle into a new equilibrium. The same tension is not registered as significantly in the velocity dispersion distribution. It seems plausible that the recent major merger may have affected the pseudo-entropy in a much stronger way than it has on the velocity dispersion. If this is the case, we expect in the near future that the system will virialize and reduce the scatter with the scaling relation. Therefore, this reasoning advocates that entropy (or pseudo-entropy) has the potential of being a good estimator for detecting recent major mergers. After removing this outlier, the logarithmic scatter is further reduced to $\sigma_{S_\mathrm{vir}|M_\mathrm{vir}}=0.12\pm0.02$, with  $\alpha = 1.72\pm 0.03$, which in turn corresponds to a lower fractional uncertainty in mass at fixed observable \Svir equal to $\sigma_{M_\mathrm{vir}|S_\mathrm{vir}} = 0.067\pm0.003$. 
    
    Inferring \Svir from observations relies on integrating the pseudo-entropy profile within the virial radius, whose knowledge is equivalent to that of the virial mass. Therefore, the use of \Svir as a mass proxy may be plagued by a circularity in the argument. The issue can be addressed via an iterative procedure, which is similar in spirit to that described by \citet{kravtsov2006new} for estimating cluster masses from the $Y_X$ mass proxy. This quantity is defined as the product of gas mass and core-excised ICM temperature, both estimated within $R_{500}$ from X-ray observations. In fact, the procedure allows to estimate the mass \mvir when one does not know a priori \rvir. Our approach would require relying on the velocity dispersion \svir in place of the X-ray temperature, to make a first rough estimate of the virial radius through a \mvir -- \svir relation. Relying then on a pre-calibrated \mvir --\svir relation (e.g. from  high-quality observations of a selected cluster sample and/or from simulations), one can then compute \rvir. The procedure can then be iterated until convergence. While exploiting the potential of \svir as a mass proxy goes beyond the scope of this paper, we plan to address this issue in a future analysis.
    
    In conclusion, we have shown that pseudo-entropy is not only a faithful tracer of the phase-space structure of a halo but also a potentially useful proxy of its total mass, thus making it an interesting tool for both dynamical studies of galaxy clusters and their cosmological application.

    \section{Pseudo-entropy profiles traced by substructures}\label{sec:S-sub}
\begin{figure}
        \includegraphics[width=8.2cm,scale=0.7,angle=0.0]{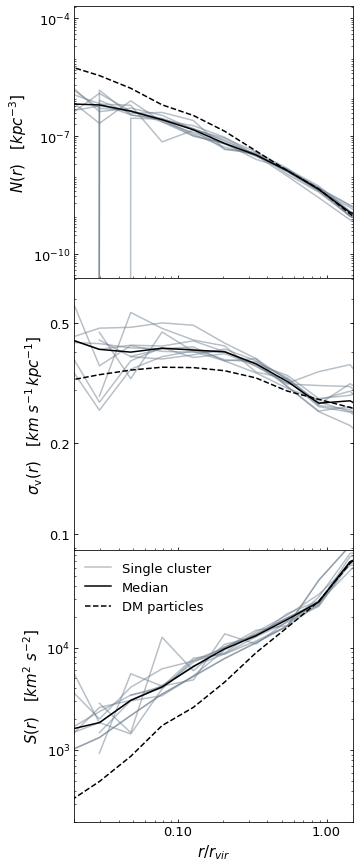}
        \caption{From top to bottom panel: number density, velocity dispersion, and pseudo-entropy profiles as traced by the substructures in the Hydro-10x simulation at $z=0$. The grey lines show the single cluster profiles, whereas the black ones reproduce the median profiles. We also report with the dashed black lines the DM particles profiles (as seen in Fig. \ref{fig:Fake_Faltenbacher-bis}) normalized to match the number density profile at $0.5\, r_\mathrm{vir}$.}
        \label{fig:Fake_Faltenbacher-SUB}
    \end{figure}

    Having established a remarkable regularity in the pseudo-entropy structure of the DM halo component, we now move to the analysis of the same quantity as traced by substructures.
    As previously discussed, observational studies \citep[e.g.,][]{biviano2013clash,biviano2016dynamics,capasso2019galaxy} demonstrated the power-law relation of the pseudo-entropy profile traced by galaxies in clusters. In our simulated clusters, {\em bona fide} galaxies correspond to gravitationally bound substructures, which we identify through the SubFind algorithm (see Sec. \ref{sec-sim}). 
    We estimate the phase-space halo structure, as traced by such substructures, from their number density profiles $N(r)$ and velocity dispersion profiles $\sigma_\mathrm{v}(r)$,
    \begin{equation}
        S(r)=\dfrac{\sigma_\mathrm{v}^2(r)}{N^{\,2/3} (r)}.
    \end{equation}
    This case differs from the previous one with DM particles, since the density employed is not the mass density, but rather the substructure number density within each cluster and the velocity dispersion profiles is derived from the statistical distribution of velocities of the substructures. For this analysis, we used the full hydrodynamical set (Hydro-1x and Hydro-10x) although we show in Fig. \ref{fig:Fake_Faltenbacher-SUB} only the profiles of the 10 clusters in the Hydro-10x. The plot illustrates the resulting number density (upper panel), velocity dispersion (central panel), and pseudo-entropy (lower panel) profiles of the single clusters along with the associated median value (solid black). In each panel, we show with the dashed curve the corresponding median profile obtained for the DM particles (as seen in Fig. \ref{fig:Fake_Faltenbacher-bis}). The density (and correspondingly, the pseudo-entropy) is normalized to match the substructures number density profiles at $0.5\, r_\mathrm{vir}$: at these large radii the two distributions are very close to each other and with this normalization one can better appreciate the differences in the central region.
   
    Note that the same universal rescaling with mass and redshift discussed in Sec. \ref{sec-method} has been applied to the quantities shown in Fig. 
    \ref{fig:Fake_Faltenbacher-SUB}. Although we do not report the not-scaled $\widetilde{\sigma_\mathrm{v}}$ and $\widetilde{S}$ as we did in Fig. \ref{fig:Fake_Faltenbacher-bis}, the internal scatter within profiles is significantly reduced after applying Equations \ref{eq:sigma(z)} and \ref{eq:S(r,z)}. The fact that both the number density and velocity dispersion profiles of substructures present the same universal scaling as the mass-density and velocity dispersion profiles traced by DM particles confirms that the self-similarity of the gravity-driven internal dynamics of clusters is preserved when traced by substructures.
    
    As for the density profiles, we note that substructures trace profiles that are shallower than the NFW profile traced by DM particles. This result confirms previous findings \citep[eg.][]{saro2006properties,van2018dark,green2019tidal} which pointed out that tidal removal of mass from merging substructures makes them more fragile in the central cluster regions, thus causing the corresponding number density profiles to flatten with respect to that traced by DM. While the velocity dispersion profiles traced by DM particles and substructures look more similar than their density profiles, we still see that substructures are characterized by a generally higher velocity dispersion, an effect that is more pronounced in central regions.
    This velocity bias, that has been also pointed out in previous studies \citep[e.g., ][]{munari2013relation,diemand2004velocity,faltenbacher2005supersonic,faltenbacher2007entropy,faltenbacher2006velocity,lau2009effects,armitage2018cluster}, is due to the effect of tidal stripping which is more effective for substructures moving with lower orbital speed. 
    As a result, these structures tend to merge into the main halo. This effect turns into a selective removal of lower-velocity substructures, thereby increasing the velocity dispersion of substructures. The resulting profiles of pseudo-entropy are thus shallower than those of DM particles, an effect that is mainly driven by the change in the density profiles.


    The best-fitting parameters describing the power-law shape of the pseudo-entropy profiles are reported in Table \ref{tab:sub} for the Hydro-1x and Hydro-10x sets of simulated clusters. Confirming the visual impression from Fig. \ref{fig:Fake_Faltenbacher-SUB}, the slope $\alpha\simeq 0.9$ is shallower than the one of $S(r)$ traced by DM particles. Furthermore, these profiles are robust against resolution, both in shape and in normalization. 
     
    \begin{table}
    \caption{Best-fit results for the normalization $S_0$ and exponent $\alpha$ in the power-law fitting of the median pseudo-entropy profiles at $z=0$ traced by substructures.}
    \centering
        \begin{tabular}{  c c c   }
            \hline
            & \thead{$S_0$} & $\alpha$   \\\midrule
             \thead{Hydro-1x} & $\left(4.44^{+0.16}_{-0.16}\right) \times 10^{4} $ & $0.89^{+0.06}_{-0.06}$\\
            \\
            \thead{Hydro-10x} & $\left(4.34^{+0.15}_{-0.15}\right) \times 10^{4} $ & $0.86^{+0.03}_{-0.03}$  \\
            \hline
        \end{tabular}%
        \label{tab:sub}
    \end{table}

    \begin{figure}
            \includegraphics[width=8.2cm,angle=0.0]{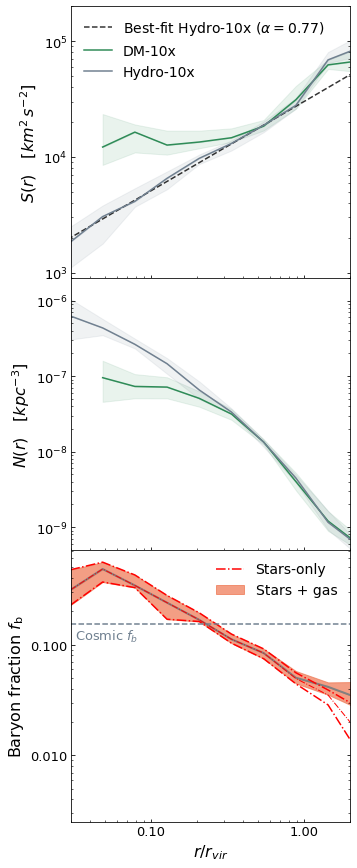}
            \caption{Comparison of profiles traced by substructures in DM-10x and Hydro-10x simulated clusters. Upper and central panels: the pseudo-entropy and number density profiles traced by the stack sample of substructures at $z=0$. Solid curves are for the median profiles, while the shaded area encompasses the 16-th and 84-th percentiles. In the upper panel, the dashed line shows the best-fit power-law relation for the pseudo-entropy profile of substructures in the Hydro-10x simulations (plotted in Fig. \ref{fig:Fake_Faltenbacher-SUB}). Lower panel: The baryon fraction profile of substructures in the Hydro-10x simulations. Green and dot-dashed red curves are the median profiles of baryon fraction when including stars and gas, and only stars, respectively. The horizontal dashed line marks the cosmic baryon fraction of the simulations ($\Omega_b/\Omega_M = 0.154$).}
            \label{fig:S-sub}
        \end{figure}
    
    \subsection{The impact of baryons on substructures}
    The results discussed so-far and presented in Fig. \ref{fig:Fake_Faltenbacher-SUB} refer to substructures identified in radiative hydrodynamical simulations. As such, they contain not only DM, but also gas and, most importantly, star particles. The latter, being originated from the dissipative collapse of gas undergoing radiative cooling, are expected to have a colder dynamics and, therefore, lower pseudo-entropy than the DM component of the subhalos \citep{dolag2009substructures}. As such, star particles are also expected to be more gravitationally bound and then more resilient against tidal disruption than the DM component. To elaborate more on this point, we compare in Fig. \ref{fig:S-sub} the profiles traced by substructures in the DM-10x and Hydro-10x sets of simulations. Having the same resolution, this comparison allows us to determine the effect of dissipative gas dynamics on the (nearly) dissipationless dynamics traced by substructures. 
    The top panel of Fig. \ref{fig:S-sub} shows that while the pseudo-entropy profiles of substructures in these two simulation sets agree in the outer cluster regions ($r>0.5 \,r_\mathrm{vir}$), they significantly differ at radii $r<0.5 \,r_\mathrm{vir}$. In particular, in DM-only simulations substructures are characterized by a plateau of pseudo-entropy in the cluster core, with no substructure found within $0.05\, r_\mathrm{vir}$. On the other hand, pseudo-entropy profiles traced by substructures in the Hydro-10x simulations are consistent with a power-law behavior over the whole $0.05 \lesssim r_\mathrm{vir} \lesssim 1$. We further investigate the origin of this difference and find that the velocity dispersion profiles of subhalos are consistent between the DM-only and Hydro runs. On the other hand, the number density profiles of substructures (central panel of Fig. \ref{fig:S-sub}) highlights the relative deficit of subhalos in the cluster core for the DM-only case. This is expected, since the presence of baryons (and in particular their stellar content which dominates the central region of subhalos) has the effect of making galaxies more gravitationally bound, therefore making them more resistant against disruption caused by the strong central tidal fields \citep{dolag2009substructures}.
    
    The deficiency of subhalos in the DM-only simulations compared to the hydrodynamical simulation could be also attributed to artificial disruption. Several authors \citep[e.g.][]{muldrew2011accuracy} argue that the halo finder may be incorrectly identifying subhalos in DM-only; while others suggest that DM simulations suffer from significant overmerging due to numerical artifacts and could be avoided by following certain criteria \citep{van2018dark}. Nonetheless, others assessed results compatible with our findings claiming the differences in the radial distribution to be the result of tidal stripping \citep[e.g.][]{weinberg2008baryon}.
 
    To further reinforce this hypothesis, we computed the baryonic fraction within substructures ($f_b$)  as a function of clustercentric distance in the Hydro-10x runs. The baryonic fraction of each substructure is defined as the ratio of the baryonic mass (which is the sum of the stellar mass $M_{\star}$, the gas mass $M_{\mathrm{gas}}$, and the black hole mass $M_{\mathrm{BH}}$) over the total mass contained in the substructure (that includes also the DM component):
    \begin{equation}\label{eq:fb}
        f_b =\frac{M_{\star} +M_{\mathrm{gas}} +M_{\mathrm{BH}}}{M_{\mathrm{TOT}}}.
    \end{equation}
    The bottom panel of Fig. \ref{fig:S-sub} describes the radial distribution of the baryon fraction within substructures. In this panel, the grey line displays the median baryon fraction in substructures, while the orange shaded area is its 68 percent dispersion within the set of simulated clusters sample. Tidal forces in the cluster center strip more easily the outer region of substructures, which is dominated by the DM component. As a result, subhalos in the central cluster region -- where tidal forces are stronger -- are characterized by a higher baryon fraction, which even exceeds the cosmic baryon fraction assumed in our simulations for $r\lesssim 0.2\, r_\mathrm{vir}$. Previous studies \citep[e.g.][]{armitage2018cluster} have already shown this effect, highlighting that galaxies selected by their stellar mass, rather than by their total mass, have a significantly lower scatter in dynamical scaling relations. Substructures in the outskirt of clusters, that have yet not felt significant effects of tidal forces, have on average baryon fractions that decrease with radial distance.
    In these regions, the baryon fraction within substructures falls well below the cosmic value assumed in the simulation ($\Omega_b/\Omega_M = 0.154$). 
    This is because the gas, that surrounds the subhalos, is ram-pressure stripped by the cluster hot atmosphere, thus leaving behind only the DM component and a minor fraction of cold star-forming gas. This hypothesis is supported by the median distribution of the baryon fraction in the cluster when accounting only for $M_{\mathrm{\star}}$ in Eq. \ref{eq:fb}, shown with the red curves in the bottom panel of Fig. \ref{fig:S-sub}: when approaching $r_\mathrm{vir}$, the baryon budget of substructures is entirely dominated by stars, while a significant contribution of gas is detected when approaching $\simeq 2\,r_\mathrm{vir}$. Therefore, substructures retain gas particles only when they are at radial distances of $r>\,r_\mathrm{vir}$, just before starting orbiting closer to the center of the cluster and being completely deprived \citep{lotz2019gone,annunziatella2016clash}.
    
    We believe that our results are not definite, and they would certainly require some more investigation to undoubtedly resolve the controversy.

    \subsection{Mass segregation in substructures}
    
\begin{figure}
        \includegraphics[width=8.2cm,angle=0.0]{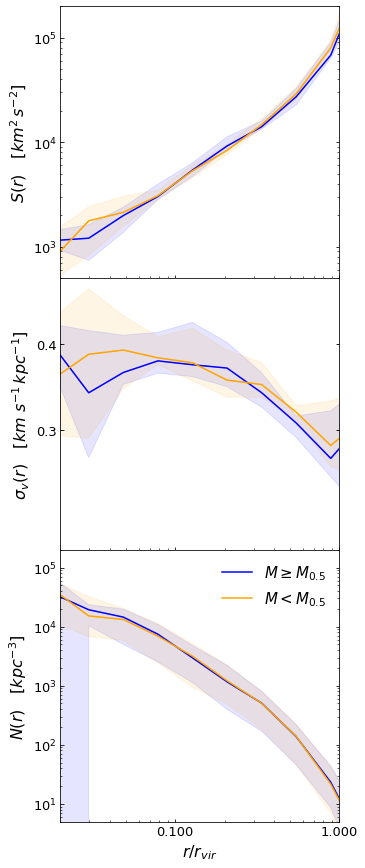} 
        \caption{Effect of mass segregation on the distribution of substructures in the Hydro-10x set of simulations. In each panel, we show with blue (orange) curves results for subhalo populations with masses larger (smaller) that the median subhalo mass found in each cluster. Shaded areas encompass the 16th-84th percentile of the distribution of profiles. From top to bottom panels we show results for pseudo-entropy, velocity dispersion and number density profiles.}
        \label{fig:mass_segreg}
    \end{figure}

    Tidal stripping is only one of the two main mechanisms responsible for the bias between DM and galaxies. As pointed out in the previous sections, we expect this phenomenon to shape the number density profile of substructures, by selectively removing lower-velocity substructures, thereby increasing their velocity dispersion and, consequently, causing a density profile shallower than that traced by DM particles. As a further test of the robustness of our results, we investigate whether the mass of the substructures introduced biases due to selection effects on the construction of the phase-space of clusters. These effects may originate from the impact of dynamical friction, which depends on the infalling mass of the orbiting substructures \citep{chandrasekhar1943dynamical}; while mass-selection biases might also be associated with the early disruption of low-mass subhalos which are more easily stripped by the strong gravitational tidal fields in the cluster central regions. To test for these effects, we split the subhalo population of each cluster within the Hydro-10x sample at their median mass $M_{0.5}$.
    We then compute the spatial and velocity radial distribution, and the pseudo-entropy profiles for these two equally populated subsamples of subhalos. Fig. \ref{fig:mass_segreg} displays the median phase-space properties (and associated 68 percent standard deviation) of the 10 clusters in the Hydro-10x divided into the high-mass group (in blue) and the low-mass group (in orange). No significant evidence of mass segregation is found in pseudo-entropy, velocity dispersion, and number density profiles, at least within the statistics allowed by our simulations. This result provides evidence that no significant mass segregation effects given by dynamical friction impacts on our results. 
    
    \vspace{.5truecm}
    
    In conclusion, the general picture emerging from the analysis of our simulations on how substructures evolve within a galaxy cluster can be summarized as follows.
    \begin{itemize}
        \item Subhalos in the outer regions are deprived of most of their diffuse gas component presumably through the ram-pressure stripping, which takes place already at distances beyond the virial radius;
        \item  During the infall within the cluster potential, substructures are stripped of their DM component, which is less resistant to tidal forces, while still preserving part of the baryonic component in the form of stars; this causes an increase of the baryonic fraction within substructures at small clustercentric radii;
        \item In the central regions, we find substructures with high baryon fractions (of about 25--45 percent) for the most part due to the presence of star particles that are gravitationally more bound, thus more resilient to tidal disruption;
        \item {These effects seem to be independent on the mass of the subhalos populating our sample of clusters because we report no evidence of mass segregation}
    \end{itemize}
    
    As a general consequence, substructures in DM-only simulations, which do not include a stellar component, are more easily destroyed and do not survive for long times in the central regions of galaxy clusters.

    \section{Pseudo-entropy profiles traced by stars}
    \label{sec:S-stars}
    
    Stars represent a collisionless component in hydrodynamical simulations, as the above discussed DM particle component. However, since they are generated from the dissipative collapse of gas particles, their phase-space structure is expected to be different from that of DM particles, whose collapse is completely non-dissipative. Therefore, we investigate the pseudo-entropy profile as traced by the stellar component as a tool to understand the different physical processes involved in its formation. 

   
    \begin{figure}
        \includegraphics[width=8.2cm,angle=0.0]{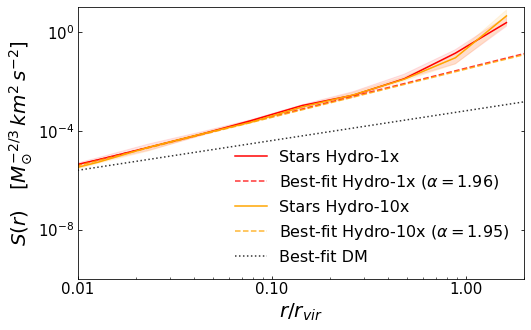}
        \caption{Pseudo-entropy profiles traced by the stars in the main halo in the Hydro-1x (solid red line) and in the Hydro-10x (solid orange line) at $z=0$. For both profiles we also report the respective intrinsic scatter of the simulated cluster samples (the shaded areas) and the best-fit profiles (dashed lines), using the same color code. To help the comparison, we report the best-fit profile for the DM particles (dotted black line) in the Hydro-1x simulation.}
        \label{fig:S-dm+s}
    \end{figure}
   
   Fig. \ref{fig:S-dm+s} illustrates the comparison between the median pseudo-entropy profiles traced by the stars belonging to the main halos of the clusters in the Hydro-1x (in red) and in the Hydro-10x (in orange) simulations. This means that here we are including only the stars belonging to the main halo, while excluding all the star particles bounded to substructures. For both profiles we also report the respective intrinsic scatter of the simulated cluster samples (the shaded areas) and the best-fit profiles (dashed lines). For reference, the dotted black line represents the best-fit curve traced by the DM particles in the Hydro-1x simulation. The stellar component recovers the power-law feature in the pseudo-entropy profiles only for $r\lesssim 0.55 \,r_\mathrm{vir}$ in both the Hydro-1x and the Hydro-10x runs, which corresponds to the clustercentric distance within which most of the stellar mass of the main halo is contained (more than 90 percent in all the clusters). In Table \ref{tab:stars} we show the best-fit parameters for both simulations: we find that the two simulations are consistent within 1$\sigma$ for the different resolutions, thus guaranteeing the convergence of our results against the resolution.

   \begin{table}
    \caption{Best-fit results for the normalization $S_0$ and exponent $\alpha$ in the power-law fitting of the median pseudo-entropy profiles at $z=0$ traced by stars.}
    \centering
        \begin{tabular}{  c c c   }
            \hline
            & \thead{$S_0$} & $\alpha$   \\\midrule
             \thead{Hydro-1x} & $\left(8.70^{+0.37}_{-0.36}\right) \times 10^{-3} $& $1.96^{+0.02}_{-0.02}$\\
             \\%
            \thead{Hydro-10x} & $\left(8.39^{+0.21}_{-0.22}\right)\times 10^{-3}$ & $1.95^{+0.02}_{-0.02}$\\
            \hline
        \end{tabular}%
        \label{tab:stars}
    \end{table}

    \begin{figure}
        \includegraphics[width=8.2cm,angle=0.0]{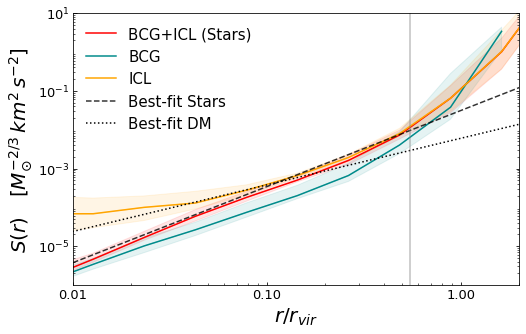}
        \caption{Pseudo-entropy profiles traced by the stars bound to the BCG (in green), ICL (in yellow) and the two combined (in red) in the Hydro-1x simulation at $z=0$. The dashed black line is the best-fit profile traced by all the stars while the dotted black line is the best-fit profile for the DM particles which has been shifted in order to match the pseudo-entropy value of the ICL profile at $0.1 \, r_\mathrm{vir}$. The vertical grey line helps the eye to visualize until which radius ($\sim 0.55 \, r_\mathrm{vir}$) the power-law holds for the sample of all the stars at this resolution.} 
        \label{fig:ICL+BCG}
    \end{figure}
    
    \subsection{ICL and BCG}
    In the hypothesis of tracing the real structure of the phase-space, one can further investigate whether the distinct dynamics and structure of the ICL and BCG composing the stars in the main halo of galaxy clusters emerge in tracing the pseudo-entropy profile. Therefore, we make use of the modified version of SubFind based on the definition of binding energy, as described in Sec. \ref{subsec:subfind}, to split the stars in ICL and BCG and we recover their single pseudo-entropy profiles. 

    The decomposition assigns on average 65 percent of the total stellar mass in the main halo to the ICL, while the rest is concentrated in the central BCG. This definition of ICL is not fully comparable with the majority of definitions often applied in simulations and observations \citep{rudick2011quantity}. For instance, several authors have identified the ICL from observational data as the stellar component with luminosity below a limiting surface brightness \citep[e.g.][]{zibetti2005intergalactic}, while other authors have modeled idealized galaxy profiles and subtracted them from the total stellar luminosity, taking the excess as ICL \citep[e.g.][]{gonzalez2007census}. Recent attempts to separate the diffuse component from the BCG in simulations were oriented in excluding a given central aperture \citep[e.g.][]{pillepich2018first,demaio2020growth} and orbiting substructures. Whereas these approaches are closer to the ones followed in the observational analysis, they implicitly make the assumption of spherical symmetry that could result in a considerable simplification of the problem. Accordingly this may lead to possible contamination from the two dynamical components which in turn can lead to biases in an analysis that aims at identifying differences in the phase-space structure of the two components. A detailed investigation of the different approaches used to separate the ICL and the BCG components is beyond the purpose of this paper, and we refer the reader to \citet{dolag2010dynamical} for a more comprehensive description of the method adopted in this work.

    Fig. \ref{fig:ICL+BCG} shows the median pseudo-entropy profiles in the Hydro-1x run: namely the sample of all the stars (red line), the BCG (in green), the ICL (in yellow) and best-fit power-law relation for stars (reported with the dashed black line) and DM particles (the dotted black line normalized to match the ICL profile at $0.1 \, r_\mathrm{vir}$).  
    According to Fig. \ref{fig:ICL+BCG}, the pseudo-entropy profile traced by the stars bound to the BCG follows a power-law shape within $r \lesssim  0.3\, r_\mathrm{vir}$ with a slope shallower than in the BCG+ICL case. On the contrary, the ICL significantly deviates from a power-law shape at all radii. However, their composition clearly follows a power-law marked by the dashed curve, at least out to the grey vertical line at $ 0.55\, r_\mathrm{vir}$. Seemingly, despite having distinct dynamics and formation histories, ICL and BCG combine together in the phase-space so as to form a power-law over a fairly large radial range, much like for DM particles, albeit with a different slope. 
    In the innermost region (for $r < 0.04$ \rvir) the pseudo-entropy traced by stars resembles the behaviour followed by the BCG. The dissipative collapse that generates these stars in the core strongly affects their dynamics: on average they are characterized by a \l'colder\r' dynamics, with relatively lower their velocity dispersion, that reduces the entropy and impacts the pseudo-entropy profile. However, moving away from the cluster core, the contribution of the high-velocity dispersion of the \l'thermalized\r' population of the ICL becomes more dominant. For distances $r \,> \,0.05\, r_\mathrm{vir}$, the ICL component prevails in the total profile.

    Having been stripped from merging galaxies by tidal interactions and having undergone phase mixing during the hierarchical halo assembly, the diffuse stellar component is often found to be closely mapping the distribution of the DM particles \citep[e.g.][]{montes2018intracluster}. However, \citet{alonso2020intracluster,contini2020mass,sampaio2020diffuse} have demonstrated that, even if this seems plausible, their radial profiles differ substantially. In this regard, we expect to observe these differences to emerge also in their phase-space structure. By plotting the best-fit profile of the pseudo-entropy traced by the DM particles, we provide support to this claim: not only in our simulation we find a significant tension in the pseudo-entropy traced by DM particles and ICL, but these differences are significant in their density and velocity dispersion profiles too (which we do not show).
   
   In conclusion, the remarkable picture arising is the universal power-law behaviour given by the composition of two dynamically distinct stellar components, which have different formation histories and are characterized by different, non-power-law, pseudo-entropy profiles.
   
    \section{Comparison with observational results}

    \label{sec:data}
    \begin{figure*}
        \includegraphics[width=16.2cm,scale=0.7,angle=0.0]{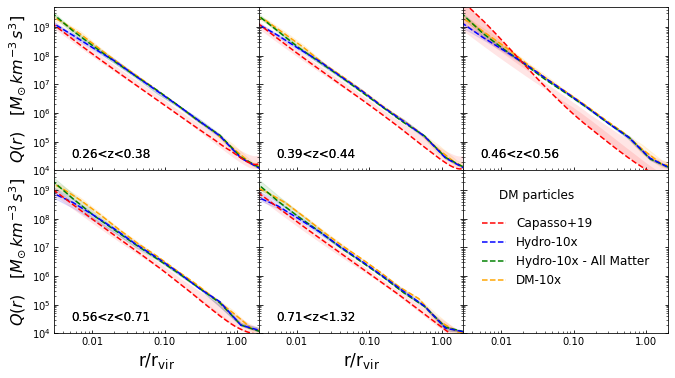}
        \caption{Phase-space density profiles as traced by DM particles within the Hydro-10x (in blue), DM-10x (in orange) and from the observed dataset from \protect\cite{capasso2019galaxy} (in red). We also report the phase-space density traced by both DM particles and baryons (stars and gas) within the Hydro-10x simulations (in green). The profiles are shown for the same 5 different redshift bins withing which the analysis of observational data have been carried out. Profiles from DM particles in the Hydro-10x have been rescaled by $(1-f_b)$ ($f_b$: cosmic baryon fraction assumed in the simulations). For simulations, lines show the median profiles while shaded areas encompass the 16-th to 84-th percentiles.}
        \label{fig:cap-pa+dm+hyd}
    \end{figure*}

    \begin{figure*}
        \includegraphics[width=16.2cm,scale=0.7,angle=0.0]{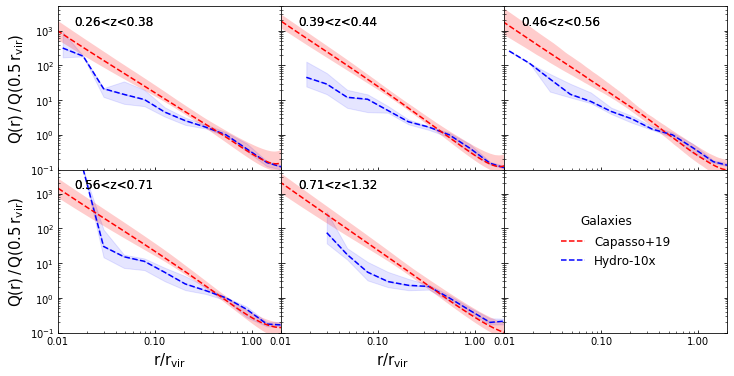}
        \caption{Phase-space density profiles as traced by substructures within the Hydro-10x simulations (in blue) and galaxies in the observational analysis in \protect\cite{capasso2019galaxy}. For each redshift interval, profiles have been normalized so as to match at $0.5$ \rvir. Lines and shaded areas have the same meaning as in Fig. \ref{fig:cap-pa+dm+hyd}.}
        \label{fig:cap-S}
    \end{figure*}
    
    Comparing our findings with observational data is fundamental to understand the capability of our simulations to correctly describe the dynamical processes leading to the formation and evolution of galaxy clusters and, ultimately, their predictive power. 
    We compare our results from the DM-10x and Hydro-10x simulations with those in \citet{capasso2019galaxy}. \citet{capasso2019galaxy} carried out an analysis of the phase-space density of clusters selected using the Sunyaev-Zel'dovich (SZ) effect in the $2500\, \mathrm{deg^2}$ South Pole Telescope (SPT)-SZ survey in the redshift range $\mathrm{0.2<z<1.3}$. The reconstruction of the phase-space of these objects is performed with MAMPOSSt \citep{mamon2013mamposst} which, adopting parametric expressions for the mass and velocity anisotropy profiles, solves the Jeans equation in spherical symmetry and recovers the 3D velocity dispersion and tracers distribution. The central 50 kpc region is excluded from their analysis, because it is identified as the cluster region affected by the presence of the BCGs. In Appendix \ref{appendixB} we discuss more in details the limits built in the method used to derive the observed pseudo-entropy profiles. We see that these may introduce potential biases and underestimate the impact of mergers.

    \subsection{Q from phase-space density}
    From the recovered mass density and velocity dispersion profiles, \citet{capasso2019galaxy} derived the profile of phase-space density, $Q(r)$, by combining results from clusters within different redshift intervals. The results of this observational analysis are shown with the red dashed lines in Fig. \ref{fig:cap-pa+dm+hyd}. The 5 different redshifts at which simulation results are shown have been chosen by selecting a snapshot at a redshift within the interval reported in each panel.
    
    In the same figure, we also report with the orange and blue curves the median profiles traced by the DM particles of the DM-10x and Hydro-10x simulated clusters, respectively. For the latter, the profiles have been rescaled by a factor $(1-f_b)^{-1}$ to take into account the baryon fraction. Finally, the green curves report the median profiles of the total mass density in the Hydro-10x simulations, i.e. adding to the DM also the contribution from the baryonic components in gas and stars.
    
    Quite remarkably, simulations and observations produce profiles with a similar slope for all the redshift ranges. The only exception is represented by the redshift interval $\mathrm{0.46<z<0.56}$, for which the observed phase-space density profile is significantly different not only from the simulated ones but also from the observed ones at the other redshifts.
    Most likely this is due to the reconstructed velocity dispersion profile in this redshift bin, which is quite irregular in comparison with that recovered in the other redshift intervals. Despite such good agreement in slope, we note a slight systematic offset in normalization between observed and simulated profiles that amounts to about 20-30 percent. While the origin of this difference is not clear, it is worth reminding that the $Q(r)$ profiles from simulations have not been obtained by reproducing as close as possible the observational procedure based on the application of the Jeans equations and the deprojection of observed profiles of number density of tracers and line-of-sight velocity dispersion. Nevertheless, we regard as quite relevant that a power-law shape of $Q(r)$ is consistently produced by observational data and simulations, when using matter density in the definition of phase-space density. Finally, we note that profiles traced by DM particles tend to slightly flatten in the innermost regions, $r\lesssim 3\times 10^{-2}\, r_\mathrm{vir}$, even if in this region the observational results are extrapolated. On the other hand, including in the analysis of the Hydro-10x simulations also the contribution of the baryonic (mostly stellar) components preserves a power-law profile extending to the innermost regions resolved in the simulations.   
      
    \subsection{Q from galaxy number density}
    As a further term of comparison, we also use the results by \citet{capasso2019galaxy} on the profiles of $Q(r)$ obtained by using the number density of tracers (galaxies), instead of the mass density profiles. Their results are shown with the red curves in Fig. \ref{fig:cap-S}, while the blue curves show the $Q(r)$ profiles obtained from simulations using substructures as tracers of halo phase-space. Owing to the difficulty of properly normalizing the number density profiles of the simulated substructures and observed galaxies, we arbitrarily fixed the normalization in such a way that simulated and observed profiles match at $0.5\, r_\mathrm{vir}$. Quite remarkably, pseudo-entropy profiles from simulations are shallower than in observations if substructures/galaxies are used to trace the phase-space structure of clusters at $r\lesssim \, 0.4 \,r_\mathrm{vir}$, while the two profiles recover the same slope at larger radii \citep[][accepted]{meneghetti2020}. This result is in line with previous findings from simulations, indicating that the observed phase space traced by cluster galaxies is not accurately described by substructures identified in simulated clusters \citep{weinmann2012fundamental,nierenberg2016missing,hirschmann2016galaxy}.

    Thus, while the slopes of phase-space density profiles traced by the total mass density are quite consistent in observational data and in simulations, the latter tend to produce lower values of phase-space density traced by substructures than in observational data. Simulations at higher resolution are needed to understand whether this is due to resolution effects, which could make substructures in simulations exceedingly fragile against the action of the tidal field in central cluster regions.

    \section{Summary}
    \label{sec-summary}
    In this paper, we presented an extensive analysis of the phase-space structure of simulated galaxy clusters. In particular, we studied the pseudo-entropy profiles $S(r)$, or equivalently the phase-space density $Q(r)$, traced by different collisionless components: dark matter (DM) particles, substructures and star particles. The analysis is based on the Dianoga set of cosmological simulations of galaxy clusters \citep{bassini2020DIANOGA}, that have been carried out with the GADGET-3 code at different resolutions and including different physics: pure N-body and hydrodynamical simulations with radiative cooling, star formation and stellar feedback models implemented following \cite{springel2003cosmological}, metal enrichment and chemical evolution following the formulation described in \cite{tornatore2007chemical}; AGN feedback as described by \cite{ragone2013brightest}. Our analysis aimed at investigating the mechanisms involved in building the phase-space structure of galaxy clusters, and in comparing predictions of simulations to observational data. We note that our analysis provides for the first time an analysis of the pseudo-entropy profiles traced by substructures, which should correspond to galaxies, and stars within the main halo of galaxy clusters.
    
    The main results can be summarized as follows:
    \begin{itemize}
        \item Pseudo-entropy profiles from simulations, as traced by all the three collisionless components (substructures, DM, and star particles), are always close to power-laws for radii smaller than the virial radius. Furthermore, these profiles scale self-similarly with mass and redshift, at least from $z=2$ which is the largest redshift we analyzed. Substructures present a profile shallower than that outlined by DM particles as an effect of the change in the density profile. Stars, on the other hand, show a steeper pseudo-entropy profile which is a consequence of the dissipative collapse of gas by radiative cooling leading to star formation.
        
        \item Stars in the main halo (i.e. not belonging to substructures) have been separated into two dynamically distinct components: those bound to the brightest cluster galaxies (BCGs) and those belonging to the intracluster light (ICL). These two components have been shown to be characterized by different pseudo-entropy profiles, e.g. BCG stars, that dominate in central regions, have a steeper slope for $S(r)$, thus turning into a lower level of pseudo-entropy at small radii. Quite remarkably, while the only BCG $S(r)$ profile is a power-law out to $0.3$ \rvir, they combine to provide an accurate power-law for the $S(r)$ profiles of the total stellar component of the main halo, extending at least out to $0.55 $ \rvir. We verified the BCG to be responsible for the shape of the total profile in the core regions ($r\leq0.04\, r_\mathrm{vir}$): stars here originate from the dissipative collapse of gas, lowering their velocity dispersion and reducing the entropy. For $r>0.55\, r_\mathrm{vir}$ the total profile follows the profile traced by the ICL, formed by the stripping of the stellar matter from satellite galaxies.
        
        \item The pseudo-entropy of DM particles integrated within the virial radius provides an accurate proxy for the total mass of galaxy clusters, with an intrinsic scatter at fixed mass of $\sigma_{M_\mathrm{vir}|S_\mathrm{vir}} = 0.067\pm0.003$. This is even smaller, by about a factor of two, than that associated to the velocity dispersion \svir  ($\sigma_{M_\mathrm{vir}|\sigma_\mathrm{v,vir}} = 0.132\pm0.003$), which is considered the tightest proxy of cluster mass. The predicted scaling follows $S_\mathrm{vir}=S_\mathrm{DM} (h(z)M_\mathrm{vir} 10^{-15} M_{\odot}^{-1})^{\gamma '}$ where $h(z)$ is the Hubble parameter in units of $100 \, km\, s^{-1}$ and $\gamma' = 1.74\pm 0.05$.
        
        \item Several factors contribute to affecting the phase-space structure of clusters, resulting in relative deviations in the power-law feature. More in detail, we found the presence of baryons to cause modest differences in the pseudo-entropy profiles traced by DM particles between the hydrodynamical and DM-only runs, but most significantly it intervened in the phase-space distribution of substructures within clusters (leading to a flattening of the pseudo-entropy profile in the central regions of clusters evolved in DM-only simulations). Furthermore, as discussed in Appendix \ref{appendixB}, the dynamical state of the single cluster can also impact the phase-space structure of these objects. This was shown to introduce a non-trivial bias in the analysis of observational data when estimates of the mass profiles were made through the resolution of the Jeans equation. The resulting pseudo-entropy profile in disturbed objects appeared to deviate in both normalization and slope from the true profile. 
        
        \item The comparisons with the observed phase-space density profiles described in \citet{capasso2019galaxy} offered the opportunity to assess the capability of simulations to predict the phase-space structure of galaxy clusters, and which evolutionary processes are responsible for it. The phase-space structure traced by DM particles is generally in good agreement with observed clusters out to the highest redshift, $z\simeq 1.3$, at which these studies have been carried out so far. On the other hand, the profiles traced by the real galaxies (in Fig. \ref{fig:cap-S}) are found to be significantly steeper than those constructed by the substructures in simulated clusters. This result is specifically evident in the central regions at all redshifts and it establishes the existing limits in cosmological simulations in reproducing the phase-space traced by satellite galaxies in clusters, a well-known problem in the literature.
    \end{itemize}
One of the general conclusions of our analysis is that pseudo-entropy profiles provide an important characterization of the phase-space structure of cluster-size halos: despite being defined from the combination of density and velocity dispersion profiles, each having a non power-law shape, $S(r)$ as traced by DM particles has a shape which is accurately described by a power-law over a fairly wide range of scales and redshift, with a normalization that scales self-similarly. Quite interestingly, also the stellar halo component developes a power-law shape of the pseudo-entropy profiles, which extends over a fairly large radial range, despite the fact that only the BCG $S(r)$ tends to form a power-law, albeit over a narrower radial range. Different shapes for $S(r)$ of DM and stars are understood in terms of the different nature of the gravitational collapse determining their respective evolution, non-dissipative for the former and dissipative for the latter. These results lend support to the idea that pseudo-entropy is a fundamental quantity, possibly more fundamental than density profiles, to characterize the non-linear evolution of a collisionless self-gravitating fluid, leading to the formation of galaxy clusters. This is also reinforced by the tiny scatter that pseudo-entropy has in the scaling relation against the total halo mass, thus possibly promoting it also to the role of precise mass-proxy for cosmological applications of galaxy clusters.

Quite interestingly, our comparison with observational data shows that a good agreement is attained only when using total density to trace pseudo-entropy. In this respect, substructures in simulations appear to trace a pseudo-entropy level in central regions, which is higher than that traced by galaxies in observational data. This is caused by the tidal disruption, that causes substructures with relatively lower orbital velocity to become fragile in central regions. As a result, only a relatively small number of substructures, with relatively high orbital velocities, survive, thus causing an excess of the pseudo-entropy with respect to what observed. A detailed analysis, also based on higher resolution simulations, will be required to assess whether this disagreement is merely due to numerical limitations of our simulations, or it is rather indicating a more fundamental lack of understanding of the processes determining the evolution of substructures inside massive cosmological halos.

    \vspace{0.3truecm}

\noindent    {\bf{ACKNOWLEDGEMENTS}}\\
We would like to thank Raffaella Capasso for providing us the observational results on the phase-space density profiles shown in Fig. \ref{fig:cap-pa+dm+hyd} and \ref{fig:cap-S}, and Andrea Biviano for useful discussions on the use of the Jeans equation to recover mass density profiles. SB, AS and IM acknowledge financial support from the PRIN-MIUR 2015W7KAWC grant, the INFN INDARK grant. AS is supported by the ERC-StG ‘ClustersXCosmo’ grant agreement 716762, and by the FARE-MIUR grant 'ClustersXEuclid' R165SBKTMA. The simulations presented in this paper have been carried out: at CINECA, with computing time provided through an ISCRA-B project, CINECA-INAF and CINECA-UNITS agreements; at the computing centre of INAF-Osservatorio Astronomico di Trieste, under the coordination of the CHIPP project \citep{bertocco2019inaf,taffoni2020chipp}; at the Tianhe-2 platform of the Guangzhou Supercomputer Center by the support from the National Key Program for Science and Technology Research and Development (2017YFB0203300). NRN acknowledges financial support from the “One hundred top talent program of Sun Yat-sen University” grant No.71000-18841229. YW acknowledges the financial support from the NSFC grant No.11803095. KD acknowledges support by the Deutsche Forschungsgemeinschaft (DFG, German Research Foundation) under Germany’s Excellence Strategy – EXC-2094 – 3907833. ER and SB acknowledge funding under the agreement ASI-INAF N.2017-14-H.0

\section*{Data Availability}
The data underlying this article will be shared on reasonable request to the corresponding author.







\appendix

\section{Normalization/scaling}\label{appendixA}
    In this appendix we provide a sketch of the derivation of the self-similar scaling with mass and redshift for the normalization of the velocity dispersion and pseudo-entropy profiles. Let us consider \rvir as the radius of a sphere within which the mean density is $\Delta_{{\mathrm{vir}}}(z)$ \citep{bryan1998statistical} times the critical density $\rho_c(z)= 3H^2(z)/8\pi G$ at that redshift. The mass \mvir enclosed in the spherical volume is given by:
    \begin{equation}
        M_{\mathrm{vir}}(z)=\frac{4}{3}\pi r^3_{\mathrm{vir}}\left[\Delta_{\mathrm{vir}}(z)\rho_c(z)\right].
    \end{equation}
    Since $\rho_{\mathrm{vir}}(z)= \Delta_{\mathrm{vir}}(z)\rho_c(z)$ and the redshift-dependent Hubble constant reads
    \begin{equation*}
        H(z)= 100 \,h\, E(z)\, km\, s^{-1} Mpc
    \end{equation*}
    where $E^2(z)=[\Omega_M(1+z)^3+\Omega_{\Lambda}]$ for a flat $\Lambda$CDM cosmology, we can explicitly write
    \begin{equation}\label{eq:rho(z)}
        \rho_{\mathrm{vir}}(z)=  \frac{\Delta_{\mathrm{vir}}(z)}{\Delta_{\mathrm{vir}}(0)}\,E^2(z)\rho_{\mathrm{vir}}(0).
    \end{equation}
    For an isothermal density profile, it is \citep{binney1987galactic}:
    \begin{equation*}
        \rho(r)= \frac{\sigma_\mathrm{v}^2}{2\pi G r^2},
    \end{equation*}
    where $\sigma_\mathrm{v}$ indicates the one-dimensional velocity dispersion.
    The relation between $\sigma_\mathrm{v}$ and the virial radius (i.e. the virial mass) then reads:
    \begin{equation}
        \sigma_{\mathrm{v,vir}}^2 \propto r^2_{\mathrm{vir}}\, \frac{\Delta_{\mathrm{vir}}(z)}{\Delta_{\mathrm{vir}}(0)}\,E^2(z).
        \label{eq:sigSS}
    \end{equation}
    Given \autoref{eq:rho(z)} and \ref{eq:sigma(z)}, the dependence of the pseudo-entropy on the cosmology is described by
    \begin{equation}
        S_{\mathrm{vir}} (z) = \frac{\sigma^2_\mathrm{v,vir}}{\rho_{\mathrm{vir}}^{2/3}(z)}\propto r^2_{\mathrm{vir}}\, E^{2/3}(z)\left[\frac{\Delta_{\mathrm{vir}}(z)}{\Delta_{\mathrm{vir}}(0)}\right]^{1/3}.
         \label{eq:entrSS}
    \end{equation}
    Therefore, by following this prescription on the density, the velocity dispersion and the pseudo-entropy we scale the vertical axis in such a way that profiles at all redshifts should overlap as long as this simple self-similar model holds. It is worth noticing that in \autoref{eq:sigSS}, for $z=0$, the velocity dispersion is proportional to \rvir so, even when considering clusters at the same redshift, we examine $ \sigma_\mathrm{v}(r)/$\rvir instead of $\sigma_\mathrm{v}(r)$. This has also its impact on the pseudo-entropy scaling, which motivated us to examine $S(r)/ r^2_{\mathrm{vir}}$.

    We note that the assuming the singular isothermal profile allows one to compute the constant of proportionality in Equations \ref{eq:sigSS} and \ref{eq:entrSS}. However the scaling against mass and redshift provided by these equations holds more in general for halos whose structural properties (i.e. halo density, velocity dispersion and orbital anisotropy profiles) do not depend on mass and redshift \citep[see also][]{bryan1998statistical}. While this does not strictly hold for the NFW profiles, the residual mass and redshift dependencies introduce only minor deviations from the purely self-similar expectation.   

    \section{Mass reconstruction from Jeans equation}\label{appendixB}
    
    The profiles of phase-space density from observational studies that we considered in Sec. \ref{sec:data} have been obtained by using the Jeans equation for a spherical system to recover the mass density profiles of galaxy clusters \citep{wolf2010accurate,mamon2013mamposst}. However, a possible lack of dynamical equilibrium or departure from spherical symmetry could introduce biases in the recovery of such mass profiles. In addition, uncertainties in the correct modeling of the orbit anisotropy profile are also expected to affect a correct mass density reconstruction \citep{merritt1987distribution}. Therefore, one may wonder whether our comparison between observed and simulated profiles of phase-space density is affected by the assumptions underlying the application of the Jeans equation. To address this issue, we decided to reconstruct pseudo-entropy profiles in simulated clusters using the Jeans equation and compare them with the intrinsic profiles. 
   
   For a spherically symmetric system in equilibrium, the Jeans equation in spherical coordinates can be cast as
    \begin{equation}
        \frac{\differential(\nu\sigma_\mathrm{v,r}^2)}{\differential r}+ \frac{\nu}{r}\left[2\sigma_\mathrm{v,r}^2-(\sigma_\mathrm{v,\theta}^2+\sigma_\mathrm{v,\phi}^2)\right] = -\nu \frac{\differential\phi}{\differential r}\,,
    \end{equation}
    where $\nu$ is the number density profile of the tracer galaxy population, $\phi$ is the gravitational potential and $\sigma_\mathrm{v,i}$ are the components of the velocity dispersion along the three spherical coordinates $r,\theta, \phi$. After integrating this equation for the radial component of the velocity dispersion profile, one obtains 
    \begin{equation}\label{eq:sigma_r2}
        \sigma_\mathrm{v,r}^2(r) = \frac{1}{\nu(r)}\int^{\infty}_r \exp\left[2\int^s_r \beta(t) \frac{\differential t}{t}\right] \nu(s) \frac{G M(s)}{s^2}\differential s
    \end{equation}
    where $G$ is the gravitational constant, $M(r)$ is the mass enclosed within $r$ and $\beta \equiv 1-\frac{\sigma_\mathrm{v,\theta}^2+\sigma_\mathrm{v,\phi}^2}{2\sigma_\mathrm{v,r}^2}$ is the velocity anisotropy profile. From this equation, the main idea is to provide a mass modelling technique which performs a Maximum Likelihood fit of the tracers distribution $\nu(r)$, assuming parametric shapes for the gravitational potential $\phi$ (or equivalently $M(r)$, the mass profile), and the velocity anisotropy profile $\beta$. In observational analyses, this computation also involves a deprojection method to pass from the observed velocity dispersion and tracer number density profiles, to their 3D counterparts. 
    In the following, we do not address the issue of deprojection that we leave to a future analysis, while we directly start from 3D information provided by simulations, so as to focus on the assumptions entering in the Jeans equation. Since the tracer population $\nu(r)$ does not necessarily follow the mass distribution $M(r)$, we must treat the two separately in the best-fit evaluation. Owing to the accurate fit provided to the density profiles produced by simulations, the NFW profile \citep{navarro1996structure,navarro1997universal} defined by
    \begin{equation}
        \rho(r)= \frac{\rho_0}{x(1+x)^2}
    \end{equation}
    is the functional form assumed for both the mass density and tracers number density profiles. In the above equation, $x= r/r_\mathrm{s}$, where the scale radius $r_{\mathrm{s}}$ is by definition the radius at which the logarithmic slope is --2. In this way the profile is determined by $r_s$ and by the normalization $\rho_0$, which are two parameters to be fitted independently for the mass profile $M(r)$ and the tracers number density profile $\nu(r)$.
    
    As for the velocity anisotropy profile $\beta(r)$, we assume its expression to be given by the Tiret model \citep{tiret2007velocity}, which proved to provide a good description for cosmological simulations of cluster-mass halos \citep{mamon2013mamposst,mamon2010universal}:
    \begin{equation}\label{eq:Tiret}
        \beta(r) = \beta_0 + \frac{r}{r+r_\mathrm{s}} \theta_B\,,
    \end{equation}
    with $r_{\mathrm{s}}$ the scale radius of the NFW profile of the tracer distribution, $\beta_0$ a normalization and $\theta_B$ the asymptotic value of the orbit anisotropy. 
    
    For each simulated cluster, we perform the best-fit of the tracer distribution $\nu(r)$ and its anisotropy velocity $\beta(r)$ profiles from simulation data. In order to recover the mass profiles in a way similar to what is done with observational data, we apply a maximum likelihood method to recover the normalization $\rho_\mathrm{0}^\mathrm{M}$ and scale radius $r_\mathrm{s}^\mathrm{M}$ of the mass profile, so that the radial velocity dispersion profile inferred from Eq. \ref{eq:sigma_r2} matches the true one measured in simulations. For each cluster, the total velocity dispersion profiles (normalized by $\sqrt{3}$) can be computed for a given model of velocity anisotropy according to $\sigma_\mathrm{v}(r) = \sigma_\mathrm{v,r}(r)\sqrt{3-2\beta(r)}$. Having reconstructed the mass density profile from the Jeans equation and the profiles of total velocity dispersion, we finally derive the reconstructed profiles of pseudo-entropy, to be compared with the true profiles. 
    
    Furthermore, in order to verify the accuracy of this procedure based on the Jeans equation to recover pseudo-entropy profiles as a function of the dynamical state of a cluster, we divided our set of simulated halos into relaxed and disturbed. To carry out this classification, we followed the prescription in \citet{biffi2016nature} and described here in Sec. \ref{sec-sim}.
    
\begin{figure*}
        \includegraphics[width=17.2cm,angle=0.0]{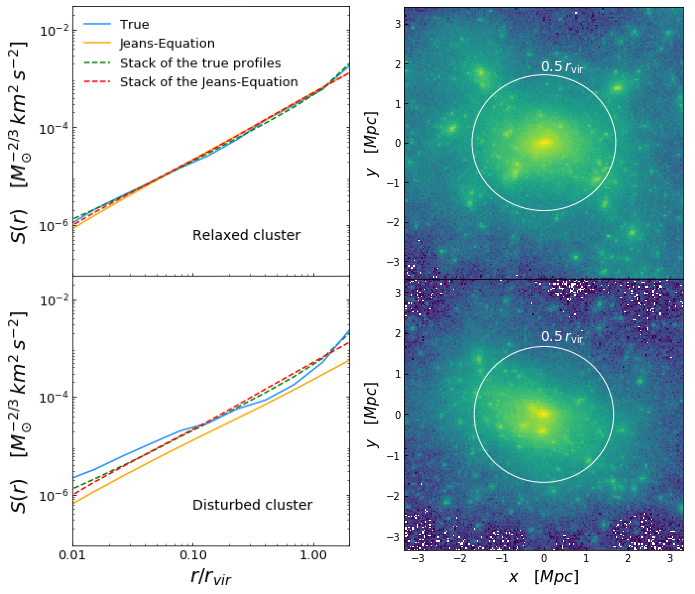} 
        \caption{On the left: pseudo-entropy profiles (DM particles) for a relaxed and a disturbed cluster in the Hydro-1x simulation at $z=0$. The dashed lines are obtained by stacking the true profiles (in cyan) or those obtained by the Jeans-Equation procedure (in orange), while the solid lines are for the profiles of the specific cluster. On the right: density maps in logarithmic scale traced by the DM particles in the same objects. The brightest colors indicate the areas at higher density. The white circles identify the 0.5 \rvir spherical region centered on the cluster center as computed by SubFind. }
        \label{fig:5bis}
    \end{figure*}

    In Fig. \ref{fig:5bis} we compare pseudo-entropy profiles recovered from the Jeans-Equation procedure with the true intrinsic profiles, for both relaxed and disturbed systems from the Hydro-1x simulation. In the right panels, we show their density maps in logarithmic scale traced by the DM particles in two clusters within the 0.5 \rvir spheres (in white). In each of the two panels on the left, we compare recovered and intrinsic profiles of $S(r)$ for both the selected cluster (solid lines), whose density map is reported on the right, and the entire cluster sample (dashed lines). Upper panels display the profiles for relaxed clusters, while lower panels are for disturbed systems. The solid lines are specific to the single cluster, being in cyan the true profile and in orange the one produced by the Jeans-Equation procedure while the dotted lines refer to the median values of all 29 clusters in the simulation set. In the upper panel, we see that true and Jeans-Equation profiles overlap at almost all radii, showing that the Jeans-Equation procedure has correctly reproduced the true profile, which is also in line with the median result. In the lower panel, we show the same profiles but when obtained for a cluster labeled as ``disturbed", since a major merger is occurring as shown by its density map. In this cluster, the true profile consistently deviates from the median and the Jeans-Equation profiles. The latter, in turn, suffers a shift in normalization but not in the slope with respect to the median profiles.

    \begin{table}
        \caption{The best-fit values obtained for the logarithmic slope in the true profile and the one recovered through the Jeans-Equation procedure for the relaxed and disturbed cluster shown in Fig. \ref{fig:5bis} with the 68 percent uncertainty. In the last row, the results obtained when fitting the median profile from the stack sample of 29 clusters in the Hydro-1x simulation at $z=0$. In the latter, the uncertainties reported are the standard deviation derived from the distribution of best-fit values for the sample.}
        \centering
            \begin{tabular}{c c c c c}
                \\
                \thead{$\alpha$}& \thead{True} & \thead{Jeans-Equation} \\\\
                \hline \\
                Relaxed & $1.203^{+0.002}_{-0.001}$ &$1.204^{+0.002}_{-0.001}$ \\\\
                Disturbed & $1.050^{+0.002}_{-0.001}$& $1.211^{+0.003}_{-0.002}$ \\\\
                Stack   &$1.222 \pm 0.002$ &$1.238 \pm 0.003$ \\\\
                \hline
            \end{tabular}
            \label{tab:slope-NCC/CC}
    \end{table}

    To quantitatively describe the differences between recovered and true pseudo-entropy profiles, we follow the same procedure described in Sec. \ref{sec-method} to fit a power-law profile, which depends on two parameters, and concentrate our interest on the accuracy in recovering the slope $\alpha$. In the first two lines of Table \ref{tab:slope-NCC/CC} we report the true and the Jeans-Equation recovered slopes for the two relaxed and the disturbed clusters shown in Fig. \ref{fig:5bis}. Consistently with the results shown in this figure, the slope is accurately recovered for the relaxed cluster, while a significant difference is found for the disturbed object. This is in line with the expectation that the procedure based on the Jeans equation, which implicitly assumes dynamical equilibrium, can introduce biases when applied to unrelaxed systems. The results from the combination of 29 simulated clusters are reported in the last row of Table \ref{tab:slope-NCC/CC}. On average, the Jeans-Equation sample seems to slightly overestimate the true value of the slope, although the difference between true and recovered slopes are relatively small once averaging over a sample of clusters with a representative mix of dynamical state.

    In conclusion, the assumption of dynamical equilibrium imposed by the Jeans equation is shown to potentially introduce a significant bias in the reconstruction the phase-space structure of unrelaxed clusters. It is worth reminding that the analysis presented here assumes the knowledge of the 3D distribution of tracers and of the velocity dispersion profiles. A full analysis aimed at including observational effects on the measurements of phase-space density of galaxy clusters would require a proper account for projection effects, which we defer to future analysis.


\bsp	
\label{lastpage}
\end{document}